\documentclass[11pt]{article}
\usepackage{jheppub}
\pdfoutput=1
\usepackage{braket} % bra and ket notation
\newcommand{\tr}{\text{tr}}

\newcommand{\mcal}{\mathcal}
\newcommand{\mbb}{\mathbb}
\newcommand{\mfrak}{\mathfrak}

\newcommand{\ra}{\rangle}

\newcommand{\wt}{\widetilde}
\newcommand{\lr}{\longleftrightarrow}
\title{Two-dimensional Anomaly, Orbifolding, and Boundary States}

\author{Ken Kikuchi, and Yang Zhou}

\affiliation{Department of Physics and Center for Field Theory and Particle Physics,\\
Fudan University, Shanghai 200433, China\\}

\abstract{
We study anomalies of discrete internal global symmetry $G$ in two-dimensional rational conformal field theories based on twisted torus partition functions. The anomaly of $G$ can be seen from the noncommutativity of two symmetry lines inserted along the nontrivial cycles of two-torus and we propose a criterion to detect the anomaly, which agrees with the truncated modular $S$-matrix approach. The obstruction for orbifolding has been recently interpreted as a mixed anomaly between $G$ and large diffeomorphisms. We clarify the relations among anomaly-free conditions, orbifoldable conditions, and invariant boundary state condition, focusing on Wess-Zumino-Witten models.
}

\preprint{}

\begin{document}
\maketitle

\section{Introduction}
Physics is constrained by symmetries. If a theory has a global symmetry, it is natural to couple it to a background gauge field, say $A$. Then the partition function of the theory becomes a functional of the background gauge field $Z[A]$. The partition function is expected to be invariant under the gauge transformation of $A$. If this is not the case, and the changes cannot be cancelled by taking local counterterms into account, the theory is said to possess an 't Hooft anomaly \cite{H80}. Since the 't Hooft anomalies are invariant under renormalization group flows, they can constrain low energy physics. See for example \cite{GKKS}, which triggered the recent development of strongly-coupled gauge theories using constraints imposed by 't Hooft anomalies. 't Hooft anomalies are also powerful in studying properties of boundaries \cite{JSY} and defects \cite{Gaiotto:2017tne}. So 't Hooft anomaly for a global symmetry $G$ can be seen as an intrinsic property of the theory independent of scales. In condensed matter physics, it has been found that 't Hooft anomalies play important roles in classifying Symmetry Protected Topological phases~\cite{CGLW,Hung:2012nf}. If $G$ is a continuous symmetry, the form of 't Hooft anomalies are tightly constrained by Wess-Zumino consistency conditions \cite{WZ71}, which implies that possible anomalies of $G$ in $d$ spacetime dimensions are classified by Chern-Simons actions in $(d+1)$ dimensions: On a $(d+1)$ manifold with a boundary, the Chern-Simons action is not gauge-invariant due to boundary terms. To cancel the boundary terms, one can couple the bulk theory to a $d$-dimensional boundary theory with 't Hooft anomaly of $G$. This is called anomaly inflow \cite{FS,CH}.\footnote{This mechanism does not always work. See for example \cite{KT} for known ``counterexamples.''} When $G$ is a finite symmetry, the anomaly inflow mechanism still works, then the 't Hooft anomalies for $G$ in $d$ dimensions are classified by $H^{d+1}(G, U(1))$ \cite{CGLW}. 

Recently, the notion of symmetry was renewed \cite{GKSW}.\footnote{See also \cite{CLSWY,BT,2group,Harlow:2018tng} for other recent trials to generalize symmetries.}  According to the modern definition, a (zero-form) global symmetry transformation in any spacetime dimensions is implemented by an invertible topological operator supported on a codimension-one defect. For instance, ordinary conserved charges in flat space (therefore the corresponding unitary symmetry transformation operators)
are defined on time slices, which have codimension one. The correlation functions of these topological operators  are invariant under smooth deformations of the defects, which is essentially the reason why they are called topological. This definition can be easily generalized to $p$-form symmetries, whose symmetry transformations are implemented by (invertible) topological operators supported on codimension-$(p+1)$ defects. Charged operators of $p$-form symmetries are $p$-dimensional. It is again natural to couple the $p$-form symmetry to background gauge fields. Charged operators of a $p$-form symmetry swipes a $(p+1)$-dimensional world-volume. So they naturally couple to $(p+1)$-form gauge fields and their gauge transformations are parametrized by $p$-form gauge connections. This is the reason for the name. By performing the gauge transformation of the background gauge fields, one can obtain 't Hooft anomalies of these generalized global symmetries.

In three-dimensional Chern-Simons theory, one-form symmetry lines and the charged operators have the same spacetime dimension one. They are realized as Wilson lines and the symmetry action is defined by linking two Wilson loops. As examples, $U(1)$ Chern-Simons theory at level $k$ and $SU(N)$ Chern-Simons theory at level $k$ both have one-form symmetries. The former has one-form symmetry $\mbb Z_k$ and the latter has 1-form symmetry $\mbb Z_N$ which coincides with the center of the gauge group~\cite{GKSW}. The 't Hooft anomaly of three-dimensional one-form symmetry $G$ can be detected by examining if the symmetry lines of $G$ are charged under themselves, namely whether the corresponding Wilson loops can be unlinked freely. In the light of the correspondence between three-dimensional Chern-Simons theories (CS) and two-dimensional rational conformal field theories (RCFT), one could ask where are these one-form 't Hooft anomalies in two--dimensional conformal field theories? One-form symmetry in three-dimensional CS corresponds to zero-form symmetry in two-dimensional RCFT, because one-dimensional symmetry lines become codimension one topological defect lines (of zero-form symmetry) in two dimensions. It is known that the expectation value of two linked Wilson loops in CS theories (in $\mbb S^3$) is given by the matrix element of the modular $S$-matrix, which can be used to detect three-dimensional one-form anomalies. Therefore it is expected that the two-dimensional anomalies (of zero-form symmetries) are also encoded in the modular $S$-matrix.~\footnote{Indeed it has been observed that the truncated modular $S$-matrix approach~\cite{HWZ} reproduces the two-dimensional zero-form anomaly conditions precisely~\cite{Y19}.}

However, a conceptual question arises. There is no notion of ``linking'' in two dimensions. How can we understand the two-dimensional anomaly without a notion of ``linking'' of topological defect lines? Our intuition is that three-dimensional ``linking'' becomes two-dimensional ``ordering.''
This is rather clear by considering a two-torus boundary in a three-dimensional bulk CS theory where the RCFT is living on. Imagine there is an ordering of acting two topological defect lines (along the two cycles respectively) on the two-torus partition function. Flipping the ordering of two topological defect lines is equivalent to unlinking (or linking) them in three dimensions.  

To be concrete, 
in this paper we focus on a zero-form global symmetry $G$ and the associated anomalies in RCFTs. 
In the radial quantization of a two-dimensional CFT onto a cylinder $\mbb S^1\times\mbb R$, the $G$ transformation on the states in Hilbert space $\cal{H}$ can be implemented by inserting a topological defect line associated to $h\in G$ along $\mbb S^1$ at a fixed time. When the $G$ line is inserted along $\mbb R$, this effectively twists the boundary condition along $\mbb S^1$ and therefore modifies the original Hilbert space to the so-called defect Hilbert space ${\cal{H}}_h$. In both insertions, the fusion of topological defect lines obey group multiplication rules. Placing the theory on a two-torus, these correspond to insert $G$ lines along two circles of the torus, which gives us the twisted torus partition function $Z_{(h,h')}$.~\footnote{Notice that since the background gauge fields in this case are one-form connections, inserting $G$ lines ($h$ and $h'$) along two cycles is equivalent to turning on background gauge fields along two cycles, with holonomy $h,h'\in G$.} Our intuition is that anomaly is the non-commutativity of insertions of two symmetry lines associated to $h$ and $h'$. To measure the non-commutativity we employ the modular transformations since they usually give important constraints in two-dimensional CFTs. We propose that the modular $S$-transformation acting on the twisted partition function can be used to detect the anomaly of $G$. Since the modular $S$-transformation exchanges two nontrivial cycles, it is naively expected that
\begin{equation}
SZ_{(h, h')}=Z_{(h', h)}\ .\label{criteria0} 
\end{equation}
If this equality is satisfied, there is no anomaly. If not, we claim there is an anomaly.~\footnote{After our main results were obtained, \cite{LS} appeared in arXiv, where they discussed the case of $G=\mbb Z_2$ using the same approach. See also~\cite{JW} for related discussion.} 

\begin{figure}[t]
\centering{}\includegraphics[scale=0.4]{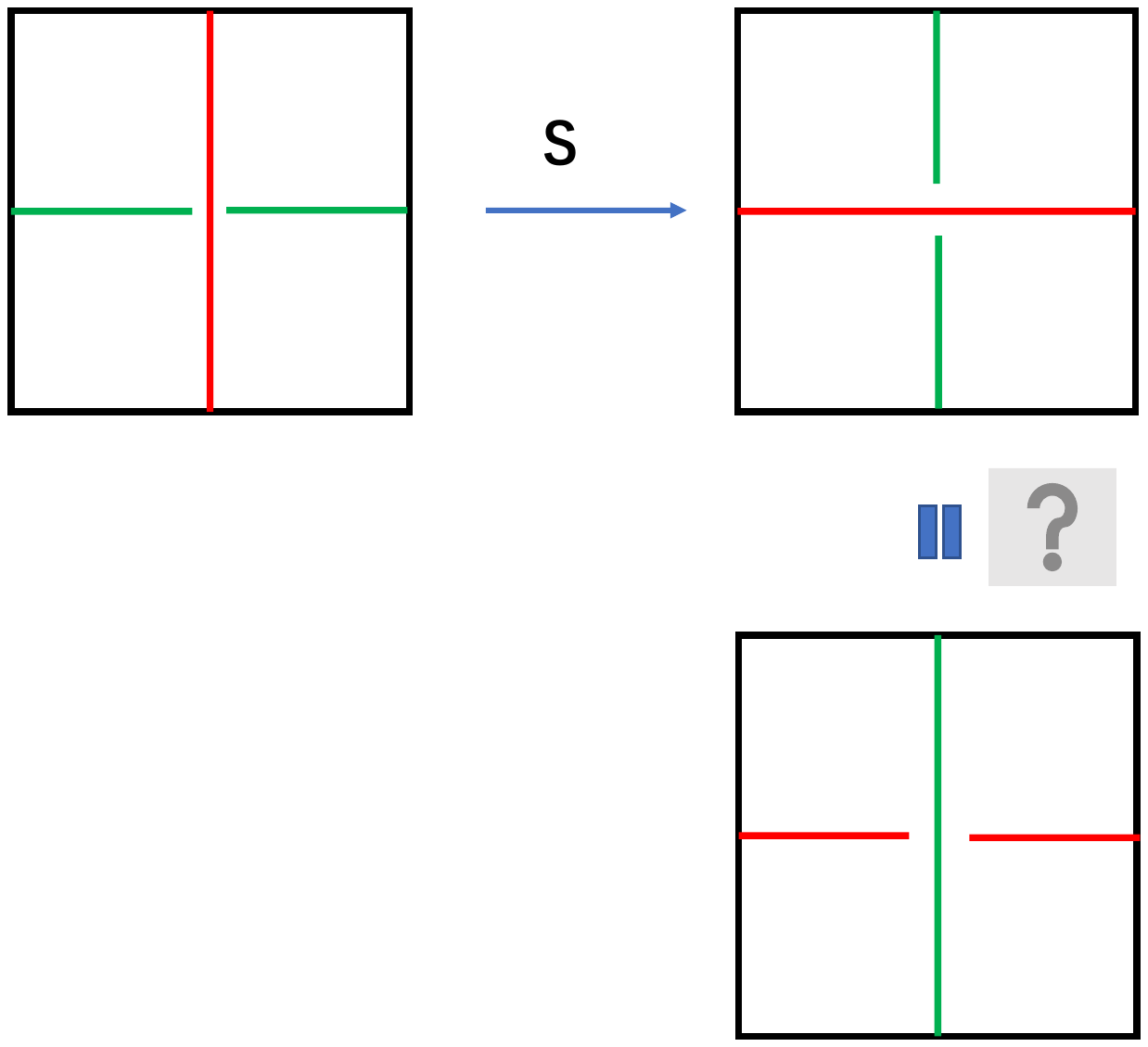}\caption{\label{Fig0} Anomaly detecting and modular transformation. %The green line and red line are $h$ and $h'$ respectively.}
We ignored the directions of the symmetry lines for simplicity.}
\end{figure}

Our criterion (\ref{criteria0}) was motivated by assuming that $Z_{(h, h')}$ is defined with a certain ordering of left insertion $h$ and right insertion $h'$, then the exchange of $h$ and $h'$ while maintaining the ordering effectively flips the ordering of $h$ and $h'$ (compared with $Z_{(h', h)}$ defined in the same manner). This is illustrated in Fig.~\ref{Fig0} where we ignored the directions of the symmetry lines. We interpret our anomaly detected by (\ref{criteria0}) as a mixed anomaly between the discrete internal global symmetry $G$ which can be read from the modular $S$-matrix and its ``$S$-dual."~\footnote{To make the terminology clear, this type of mixed 't Hooft anomaly is the one we focus in this paper. This may be distinguished from the usual ``F-symbol'' 't Hooft anomaly in $2d$. We thank Yuji Tachikawa for explanation.}

Using the twisted partition function we also study the relation between anomaly-free condition and the orbifolding condition. The obstruction for orbifolding was recently interpreted as the mixed anomaly between $G$ and large diffeomorphisms~\cite{NY}. A slightly generalized version of the orbifolding criterion in~\cite{NY} can be written as \begin{equation}\label{criteria2} Z_{(h^n, h')}=Z_{(1,h')}\ ,\end{equation} where $h$ is the generator of a cyclic subgroup $\mbb Z_n$. When $G$ is a product group including many subgroups with generators $h_i$, the orbifolding condition (\ref{criteria2}) has to be imposed for each subgroup.\footnote{One may worry that both equation (\ref{criteria0}) and equation (\ref{criteria2}) involve many group elements $h,h'\in G$, but it turns out for cyclic subgroups one only needs to consider generators $h,h'$.} We explicitly show the difference between our anomaly-free condition and obstruction of orbifolding focusing on Wess-Zumino-Witten (WZW) models.

Yet another approach to detect our anomaly is by looking at the boundary conformal field theories. In the anomaly inflow mechanism, the anomalous boundary theory cannot enjoy a boundary intuitively because $\partial^2=0$. Consequences of this observation were studied, say, in \cite{JSY} by high energy physicists and in \cite{HTHR} by condensed matter physicists. A theory is called edgeable if there is no obstruction to assign a boundary state while maintaining the symmetry. This can be written as
\[ G\text{-anomalous}\Rightarrow\text{unedgeable}, \]
or equivalently
\begin{equation}
    \text{edgeable}\Rightarrow G\text{-anomaly free}.\label{edgecohomology'}
\end{equation}
Recently it was conjectured in \cite{Y19} that if a $G$-invariant boundary state exists, then $G$ does not have anomaly nor mixed anomalies with other internal symmetries. In this paper we support this conjecture by providing further evidences.

\begin{figure}[t]
\centering{}\includegraphics[scale=0.4]{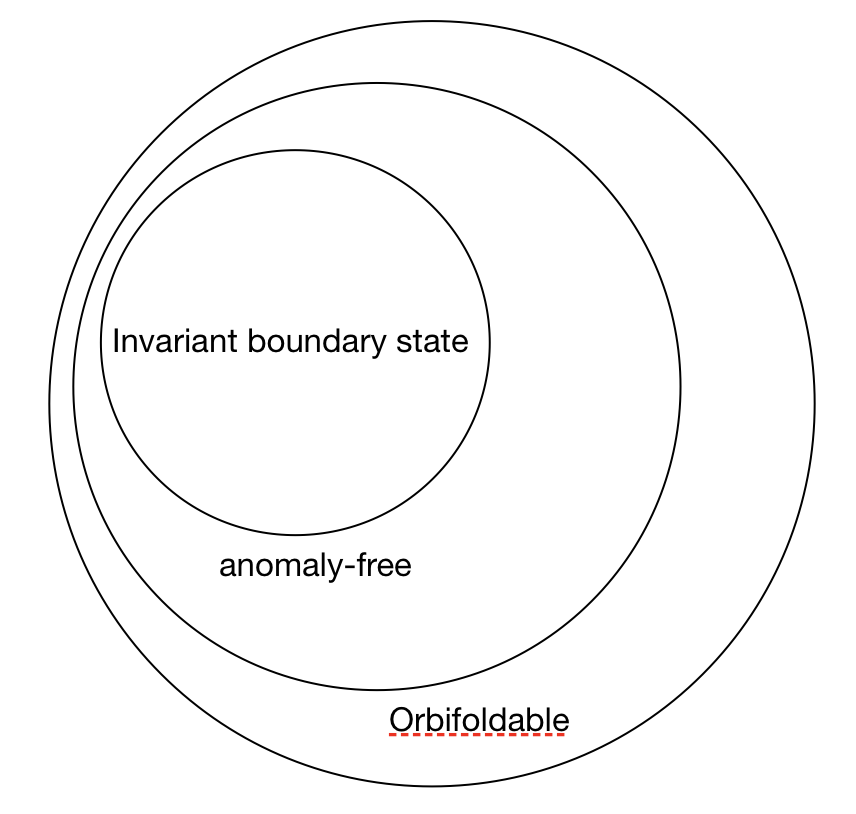}\caption{\label{Fig1} $G$-invariant boundary state condition, anomaly-free condition, and $G$-orbifolding condition.}
\end{figure}

As a byproduct, we clarify the relations among $G$-invariant boundary state condition, anomaly-free condition, and $G$-orbifolding condition, as explained in the diagram in Fig.\ref{Fig1}.

This paper is structured as follows. In section \ref{SZZ}, we propose a criterion to detect anomalies of zero-form internal discrete symmetries in two-dimensional diagonal RCFTs. We test our proposal against WZW models and some minimal models. We also explain the relations between anomaly-free conditions and orbifolding conditions. We find the former is stronger than the latter. In section \ref{IB}, we move to the boundaries. We clarify the relations between conditions which guarantee the existence of invariant boundary states and anomaly-free conditions. Our results show that when there exists an $H$-invariant boundary state, $H$ cannot have anomaly nor mixed anomalies with others.
Finally, we summarize our results and discuss future directions in section \ref{discussion}. There are 3 appendices. We review the generalized orbifolding procedure in Appendix \ref{Go}. In appendix \ref{invanom}, we complete the proof of a claim that an existence of $H$-invariant boundary state is equivalent to $H$ is ``anomaly-decoupled.'' In appendix \ref{Ch}, we complete the discussion of \cite{NY} to include the $D_{2l}$ type WZW model.

\section{Detecting our anomaly}\label{SZZ}
A modern approach to characterize a zero-form global symmetry in two-dimensional quantum field theories is by the so called invertible topological defect lines (TDLs)~\cite{Verlinde:1988sn,Petkova:2000ip,Fuchs:2002cm,CLSWY,BT,Frohlich:2004ef,Fuchs:2003id,Fuchs:2004dz,Fuchs:2004xi,Fjelstad:2005ua,Frohlich:2006ch,Runkel:2005qw,Davydov:2010rm,Frohlich:2009gb}. These are codimension one objects which implement the unitary symmetry transformations when contracted along a loop around a local operator. The fusion of the topological defect lines obey the group multiplication rules. The effect of the insertion of these topological defect lines are rather clear. In radial quantization onto a cylinder for example, they implement the symmetry transformations when inserted along the spatial circle and twist the Hilbert space when inserted along the time direction. As the simplest example, the torus partition function of free two-dimensional massless Dirac fermions can include these insertions by twisting boundary conditions along both temporal and spatial directions. In general, when the defect lines are present in both temporal and spacial directions, there appear some ambiguities originating from their crossing points. This is the general reason to cause some anomalies. Though the anomaly can be seen from the non-commutativity of insertions in temporal and spacial directions as discussed in the introduction, in general it is difficult to make a notion of the ordering of the insertions of topological defect lines. Therefore it is challenging to detect the anomaly associated to global symmetry $G$ from twisted partition functions for a given CFT$_2$. 

In this section, we propose a criterion to detect anomalies of zero-form internal symmetries by performing modular $S$-transformations on the twisted partition functions with topological defect lines inserted along both spatial and temporal directions. This was motivated by imaging that there is an ordering of the insertions of defect lines into the untwisted torus partition functions. We illustrate our proposal in detail as follows.

For simplicity we consider an Abelian global symmetry $G$. We consider the CFT on a torus with modulus $\tau$ and couple the theory to external background gauge fields. The consequence is that we have twisted boundary conditions representing the group $G$. For convenience let us denote the boundary conditions by $(h_t, h_x)$ where they correspond to set the twisted boundary conditions $h\in G$ in imaginary time direction and spatial direction respectively. With the convention that left subscript twisting the time and right subscript twisting the space, we have the twisted partition function denoted by $Z_{(h_t,h_x)}(\tau)$. In the language of topological defect lines, the torus partition functions with defect lines along the temporal direction or spatial direction are given by
\begin{equation}
Z_{(h,1)}(\tau)=\text{Tr}_\mathcal{H} [\hat h q^{L_0-c/24}\bar q^{\bar L_0-c/24}]\ ,\quad  Z_{(1,h)}(\tau)=\text{Tr}_{\mathcal{H}_h} [q^{L_0-c/24}\bar q^{\bar L_0-c/24}]\ ,\label{Zh1}
\end{equation}
where $q=\exp(2\pi i \tau)$ and $\bar q=\exp(-2\pi i \bar\tau)$. They are related by a modular $S$-transformation
\begin{equation}
SZ_{(h,1)}(\tau)=Z_{(1,h)}(\tau)\ .\label{Z1h}
\end{equation}
Under modular $T$-transformations,
\begin{equation}
T^nZ_{(1,h)}(\tau) = Z_{(h^n,h)}(\tau)\ .\label{Zhnh}
\end{equation}
For a cyclic symmetry $G$ of order $N$, there is apparently a consistency condition
\begin{equation}
Z_{(h^N,h)}(\tau) = Z_{(1,h)}(\tau)\ ,\label{ZhNh}
\end{equation} coming from the group fusion of topological defect lines. The violation of this condition has been recently interpreted as the mixed anomaly between $G$ and large diffeomorphisms~\cite{NY}. The slightly generalized version is given by (\ref{criteria2}).
We are motivated by another consistency condition
\begin{equation}\label{SZ}
SZ_{(h,h)}(\tau) = Z_{(h,h)}(\tau)\ ,
\end{equation} whose slightly generalized version is nothing but (\ref{criteria0}). We propose that violation of this condition will reflect the existence of anomaly of $G$.
When $G=\mbb Z_2$ we check that (\ref{SZ}) is the correct criterion to detect the anomaly by examining the known CFT examples. When $G=\mbb Z_N$, $h$ is the generator of $\mbb Z_N$ and we have to understand the criterion (\ref{SZ}) in a truncated version since the spectrum of the twisted partition function $Z_{(h,h)}$ in general is very rich. We test our proposal by examining many examples below.

\subsection{Examples}\label{examples}
In this subsection, we explain our criterion to detect anomalies of zero-form internal symmetries in Wess-Zumino-Witten (WZW) models. We first briefly review WZW models~\cite{CFT}.

The elementary bosonic field $g(x)$ is valued in a unitary representation of the (semisimple) group $\hat G$,\footnote{We use $\hat G$ for the group labeling a WZW model to distinguish with global symmetry $G$.} $g(x)\in \hat G$. The action is given by
\begin{equation}
    S=\frac k{8\pi}\int_Xd^2x\text{Tr}\left(\partial^\mu g^{-1}\partial_\mu g\right)+\frac k{12\pi i}\int_Bd^3y\epsilon_{\alpha\beta\gamma}\text{Tr}\left(\tilde g^{-1}\partial^\alpha\tilde g\tilde g^{-1}\partial^\beta\tilde g\tilde g^{-1}\partial^\gamma\tilde g\right),\label{WZW}
\end{equation}
$B$ is a three-dimensional manifold bounded by the two-dimensional spacetime $X$, $\partial B=X$, and $\tilde g$ is an extension of $g$ to $B$. By requiring that the extension does not cause ambiguity, $k$, which is called the level, has to be an integer, and below we will take $k$ be nonnegative without loss of generality. The theory (\ref{WZW}) is usually called $\hat G_k$ WZW model. A fact which will be used frequently below is that the primaries of the models correspond to affine weights for general Lie groups $\hat G$, labeled by a set of non-negative integers called affine Dynkin labels
\[ \hat\mu=[\mu_0;\mu_1,\cdots,\mu_r], \]
where $r$ is the rank of $\hat G$. If it labels a finite-dimensional irrep, the weights are called integral. The affine Dynkin labels are constrained by the level
\begin{equation}
    k=\mu_0+\sum_{j=1}^ra_j^\vee\mu_j,\label{WZWk}
\end{equation}
where $a_j^\vee$ are comarks (a.k.a. dual Kac labels). Thus for a finite $k$, the number of distinct affine weights, called affine dominant weights $\hat\mu\in P^k_+$, is finite, $|P^k_+|<\infty$.
In equations, the set is defined by
\begin{equation}
    P^k_+:=\left\{\hat\mu\Big|\mu_j\ge0\&0\le\sum_{j=1}^ra_j^\vee\mu_j\le k\right\}.\label{Pk+}
\end{equation}
We emphasize that in the set $P^k_+$, one has to include all possible affine weights. In other words, it obeys the ``totalitarian principle."
The affine weights can be expanded in terms of basis $\hat\omega_j$ $(j=0,\dots,r)$ called fundamental weights as
\begin{equation}
    \hat\mu=\sum_{j=0}^r\mu_j\hat\omega_j.\label{hatmu}
\end{equation}
Therefore, affine weights can be considered as vectors whose components are given by affine Dynkin labels. As usual vectors, one can define a scalar product of two affine weights $(\hat\mu,\hat\lambda)\in\mbb R$, which can be computed using the quadratic form matrix $F_{jl}=F_{lj}$
\begin{equation}
    (\hat\mu,\hat\lambda)=\sum_{j,l=1}^r\mu_j\lambda_lF_{jl}\ ,\label{inner}
\end{equation}
where the sum runs from one and not from zero since $\hat\omega_0$ has zero scalar products with any fundamental weights. The explicit values of the quadratic form matrices can be found in \cite{CFT}.

There are internal global symmetries of WZW models with affine algebra $\hat{\text{g}}$ called outer automorphisms $\mcal O (\hat{\text{g}})$, which is isomorphic to the center of the Lie group $\hat G$, $B(\hat G)$. They act on the affine weights by sending $\hat\mu$ to $A\hat\mu$ for an element of the group $A\in\mcal O(\hat{\text{g}})$. Explicit actions of course depend on $\hat G$ and will be given when we discuss $\hat G_k$ WZW models below.

The central charges and conformal weights of a primary field labeled by an affine weight $\hat\mu$ are given by
\begin{equation}
    c=\frac{k\dim\text{Lie}(\hat G)}{k+g},\quad h_{\hat\mu}=\frac{(\hat\mu,\hat\mu+2\hat\rho)}{2(k+g)},\label{ch}
\end{equation}
where $g$ is the dual Coxeter number
$ g:=\sum_{j=1}^ra_j^\vee+1$
and $\hat\rho:=\sum_{j=0}^r\hat\omega_j$ is the so called affine Weyl vector. With these knowledge, one can compute partition function of the WZW models. Since they can be seen as diagonal rational conformal field theories (RCFTs), the torus partition functions are given by
\begin{equation}
    Z=\sum_{\hat\mu\in P^k_+}\bar\chi_{\hat\mu}\chi_{\hat\mu},\label{WZWpartfunc}
\end{equation}
where $\chi_{\hat\mu}$ is the character of the conformal family associated to the primary state labeled by $\hat\mu$.

For our purpose to detect the anomalies, one can first compute $Z_{(h,1)}(\tau)$
defined in (\ref{Zh1}). By modular $S$-transformation we will get $Z_{(1,h)}(\tau)$ and then by modular $T$-transformation (multiple times) we will obtain $Z_{(h^\ell,h)}(\tau)$.

\subsubsection{$SU(2)_k$ WZW}
As a warm up, let us study the $SU(2)_k$ WZW theory defined on a two-torus. The modular $S$-matrix of the model is given by
\begin{equation}
    S_{jj'}=\sqrt{\frac2{k+2}}\sin\left(\frac{\pi(2j+1)(2j'+1)}{k+2}\right),\label{S}
\end{equation}
where $j,j'$ are spins $0\le j,j'\le\frac k2$. A small computation shows
\begin{equation}
    (-)^{2j}S_{jj'}=S_{j,\frac k2-j'}.\label{propsu2}
\end{equation}
Using this, one can compute $SZ_{(h,h)}$
\begin{align*}
    SZ_{(h,h)}(\tau)&=S\sum_{j=0,1/2,\dots,k/2}(-i)^k(-)^{2j}\chi_j(\tau)\bar\chi_{\frac k2-j}(\bar\tau)\\
    &=\sum_{j,j_1,j_2}(-i)^k(-)^{2j}S_{jj_1}S_{\frac k2-j,j_2}\chi_{j_1}(\tau)\bar\chi_{j_2}(\bar\tau)\\
    &=\sum_{j,j_1,j_2}(-i)^k(-)^{2j}S_{jj_1}(-)^{2j_2}S_{jj_2}\chi_{j_1}(\tau)\bar\chi_{j_2}(\tau)\\
    &=\sum_{j,j_1,j_2}(-i)^k(-)^{2j_2}S_{jj_1}S_{j,\frac k2-j_2}\chi_{j_1}(\tau)\bar\chi_{j_2}(\bar\tau)\\
    &=\sum_{j_1,j_2}(-i)^k(-)^{2j_2}\delta_{j_1,\frac k2-j_2}\chi_{j_1}(\tau)\bar\chi_{j_2}(\tau)\\
    &=\sum_j(-i)^k(-)^{2(\frac k2-j)}\chi_j(\tau)\bar\chi_{\frac k2-j}(\tau)\\
    &=(-)^kZ_{(h,h)}(\tau),
\end{align*}
where we used (\ref{propsu2}) twice. Thus $Z_{(h,h)}$ is invariant under the modular $S$-transformation iff $k\in2\mbb Z$, while the partition function flips sign iff $k\in2\mbb Z+1$. In fact this approach to detect $\mbb Z_2$ anomaly by computing $Z_{(h,h)}(\tau)$ and its $S$-transformation can be used for many other theories with a global $\mbb Z_2$ and we believe it is a general criterion.

When we move to discrete global symmetry larger than $\mbb Z_2$, in general one cannot get the mismatch of twisted partition functions as an overall phase, rather it appears as a unitary phase matrix. To illustrate this fact, let us study the $SU(3)_k$ WZW model which has a $\mbb Z_3$ global symmetry.

\subsubsection{$SU(3)_k$ WZW}
In this case, the outer automorphism group is $\mbb Z_3$ which is isomorphic to the center of $SU(3)$. To study the anomaly, we consider partition functions twisted by this $\mbb Z_3$. Let us first consider the case $k=1$. The relevant twisted partition function is given by
\begin{equation}\label{twistedZSU3} Z_{(h,h)}(\tau)=\omega^2\bar\chi_3(\bar\tau)\chi_1(\tau)+\bar\chi_{\bar3}(\bar\tau)\chi_3(\tau)+\omega\bar\chi_1(\bar\tau)\chi_{\bar3}(\tau), \end{equation}
where ($1=[1;0,0],3=[0;1,0],\bar3=[0;0,1]$) are three primaries and $\omega=e^{2\pi i/3}$. One can rewrite (\ref{twistedZSU3}) in a matrix form by choosing a basis $\{1,3,\bar3\}$. Therefore $Z_{(h,h)}(\tau)$ can be represented as a (special unitary) matrix
\[ Z_{(h,h)}\lr\begin{pmatrix}0&\omega^2&0\\0&0&1\\\omega&0&0\end{pmatrix}=:U\ , \]
where the rows and columns label $\chi$ and $\bar\chi$, respectively. Performing the modular $S$-transformation on the twisted partition function, one obtains
\[ SZ_{(h,h)}(\tau)=\omega^2\bar\chi_1(\bar\tau)\chi_{\bar3}(\tau)+\omega\bar\chi_3(\bar\tau)\chi_1(\tau)+\bar\chi_{\bar3}(\bar\tau)\chi_3(\tau), \]
or in the matrix form
\[ SZ_{(h,h)}\lr\begin{pmatrix}0&\omega&0\\0&0&1\\\omega^2&0&0\end{pmatrix}=:U'. \]
The mismatch between $Z_{(h,h)}$ and $SZ_{(h,h)}$ can be computed by $U^{-1} U'$:\footnote{There is a quicker way to compute this difference; see the beginning of the subsection \ref{SZZWZW}.}
\begin{equation}\label{Dmatrix} D:=U^{-1} U'=\begin{pmatrix}\omega&0&0\\0&\omega^2&0\\0&0&1\end{pmatrix}. \end{equation}
This matrix $D$ is always unitary because the modular $S$-transformation is a unitary transformation. $SZ_{(h,h)}$ equals to $Z_{(h,h)}$ iff $D$ equals to the identity matrix. Following our proposal, this means that the theory is free of our anomaly. A nontrivial $D$ matrix (\ref{Dmatrix}) in the current case of $SU(3)_1$ reflects that there is an anomaly in $\mbb Z_3$.  Notice that in the theory of $SU(3)_1$, the three WZW primaries are precisely the three topological defect lines associated to $\mbb Z_3$ global symmetry.
Later will use $\wt D$ to replace $D$ when not all the primaries are the topological defect lines.  $\wt D$ is an analogy of $D$ however truncated into the subspace supported only by the topological defect lines. 

Put into a unified fashion, our proposal is
\begin{equation}
    \text{anomaly free}\iff \wt D=I_{|G|}.\label{criteria}
\end{equation}
In the present case of $SU(3)_1$ $\wt D=D$. From (\ref{Dmatrix}), we see that $\mbb Z_3$ is anomalous.

This criterion would need a comment. Anomalies appear as phase mismatch between partition functions before and after gauge transformations of background fields (modulo local counterterms). Why we claim the anomalies appear as nondiagonal matrices? The reason is simple: modular $S$-transformations are $not$ gauge transformations of background fields. So our anomalies do not have to appear as phases.

Let us move to $SU(3)_k$ WZW model with $k=2$. The twisted partition function is given by
\[ Z_{(h,h)}=\omega\bar\chi_{[0;2,0]}\chi_{[2;0,0]}+\omega^2\bar\chi_{[0;1,1]}\chi_{[1;1,0]}+\bar\chi_{[1;1,0]}\chi_{[1;0,1]}+\bar\chi_{[0;0,2]}\chi_{[0;2,0]}+\omega\bar\chi_{[1;0,1]}\chi_{[0;1,1]}+\omega^2\bar\chi_{[2;0,0]}\chi_{[0;0,2]}. \]
In a matrix form, $Z_{(h,h)}$ can be represented as
\[ Z_{(h,h)}\lr\begin{pmatrix}0&0&0&\omega&0&0\\0&0&0&0&\omega^2&0\\0&1&0&0&0&0\\0&0&0&0&0&1\\0&0&\omega&0&0&0\\\omega^2&0&0&0&0&0\end{pmatrix}=:U \]
in the basis $\{[2;0,0],[1;1,0],[1;0,1],[0;2,0],[0;1,1],[0;0,2]\}$. Performing the modular $S$-transformation one obtains
\[ SZ_{(h,h)}=\omega^2\bar\chi_{[0;2,0]}\chi_{[2;0,0]}+\omega\bar\chi_{[0;1,1]}\chi_{[1;1,0]}+\bar\chi_{[1;1,0]}\chi_{[1;0,1]}+\bar\chi_{[0;0,2]}\chi_{[0;2,0]}+\omega^2\bar\chi_{[1;0,1]}\chi_{[0;1,1]}+\omega\bar\chi_{[2;0,0]}\chi_{[0;0,2]}, \]
or in a matrix form in the same basis as above
\[ SZ_{(h,h)}\lr\begin{pmatrix}0&0&0&\omega^2&0&0\\0&0&0&0&\omega&0\\0&1&0&0&0&0\\0&0&0&0&0&1\\0&0&\omega^2&0&0&0\\\omega&0&0&0&0&0\end{pmatrix}=:U'. \]
The $D$-matrix can be computed as
\[ D:=U^{-1} U'=\begin{pmatrix}\omega^2&0&0&0&0&0\\0&1&0&0&0&0\\0&0&\omega&0&0&0\\0&0&0&\omega&0&0\\0&0&0&0&\omega^2&0\\0&0&0&0&0&1\end{pmatrix}. \]
How should we detect the anomaly from this result?
As mentioned earlier we should truncate our matrix $D$ to $\wt D$ in the current case. This is in the same spirit as \cite{HWZ}. Truncation means that examining only those characters of primaries corresponding to topological defect lines of $\mbb Z_3$ symmetry. This would be sufficient to detect anomalies. The characters are those labeled by $\hat\mu=[2;0,0],[0;2,0],[0;0,2]$. Thus we have to truncate the $D$-matrix to $|G|\times|G|$ submatrix
\[ \wt D=\begin{pmatrix}\omega^2&0&0\\0&\omega&0\\0&0&1\end{pmatrix}. \]
Following the proposal (\ref{criteria}) the $\mbb Z_3$ of the $SU(3)_2$ has an anomaly.

Finally, let us examine the case $k=3$. The twisted partition functions are given by
\begin{align*}
Z_{(h,h)}&=\bar\chi_{[0;3,0]}\chi_{[3;0,0]}+\omega\bar\chi_{[0;2,1]}\chi_{[2;1,0]}+\omega^2\bar\chi_{[1;2,0]}\chi_{[2;0,1]}+\omega\bar\chi_{[0;1,2]}\chi_{[1;2,0]}+\bar\chi_{[1;1,1]}\chi_{[1;1,1]}\\
    &~~+\omega\bar\chi_{[2;1,0]}\chi_{[1;0,2]}+\bar\chi_{[0;0,3]}\chi_{[0;3,0]}+\omega\bar\chi_{[1;0,2]}\chi_{[0;2,1]}+\omega\bar\chi_{[2;0,1]}\chi_{[0;1,2]}+\bar\chi_{[3;0,0]}\chi_{[0;0,3]}.
\end{align*}
or
\[ Z_{(h,h)}\lr\begin{pmatrix}0&0&0&0&0&0&1&0&0&0\\0&0&0&0&0&0&0&\omega&0&0\\0&0&0&\omega^2&0&0&0&0&0&0\\0&0&0&0&0&0&0&0&\omega&0\\0&0&0&0&1&0&0&0&0&0\\0&\omega&0&0&0&0&0&0&0&0\\0&0&0&0&0&0&0&0&0&1\\0&0&0&0&0&\omega&0&0&0&0\\0&0&\omega&0&0&0&0&0&0&0\\1&0&0&0&0&0&0&0&0&0\end{pmatrix}=:U, \]
in the basis $\{[3;0,0],[2;1,0],[2;0,1],[1;2,0],[1;1,1],[1;0,2],[0;3,0],[0;2,1],[0;1,2],[0;0,3]\}$. The modular $S$-transformation of the twisted partition function gives
\begin{align*}
    SZ_{(h,h)}&=\bar\chi_{[0;3,0]}\chi_{[3;0,0]}+\omega^2\bar\chi_{[0;2,1]}\chi_{[2;1,0]}+\omega\bar\chi_{[1;2,0]}\chi_{[2;0,1]}+\omega\bar\chi_{[0;1,2]}\chi_{[1;2,0]}+\bar\chi_{[1;1,1]}\chi_{[1;1,1]}\\
    &~~+\omega^2\bar\chi_{[2;1,0]}\chi_{[1;0,2]}+\bar\chi_{[0;0,3]}\chi_{[0;3,0]}+\omega^2\bar\chi_{[1;0,2]}\chi_{[0;2,1]}+\omega\bar\chi_{[2;0,1]}\chi_{[0;1,2]}+\bar\chi_{[3;0,0]}\chi_{[0;0,3]},
\end{align*}
or
\[ SZ_{(h,h)}\lr\begin{pmatrix}0&0&0&0&0&0&1&0&0&0\\0&0&0&0&0&0&0&\omega^2&0&0\\0&0&0&\omega&0&0&0&0&0&0\\0&0&0&0&0&0&0&0&\omega&0\\0&0&0&0&1&0&0&0&0&0\\0&\omega^2&0&0&0&0&0&0&0&0\\0&0&0&0&0&0&0&0&0&1\\0&0&0&0&0&\omega^2&0&0&0&0\\0&0&\omega&0&0&0&0&0&0&0\\1&0&0&0&0&0&0&0&0&0\end{pmatrix}=:U', \]
The $D$-matrix is given by
\[ D:=U^{-1} U'=\begin{pmatrix}1&&&&&&&&&\\&\omega&&&&&&&&\\&&1&&&&&&&\\&&&\omega^2&&&&&&\\&&&&1&&&&&\\&&&&&\omega&&&&\\&&&&&&1&&&\\&&&&&&&\omega&&\\&&&&&&&&1&\\&&&&&&&&&1\end{pmatrix}. \]
Truncating the $D$-matrix to $|G|\times|G|=3\times3$ submatrix spanned by $\hat\mu=[3;0,0],[0;3,0],[0;0,3]$, one gets
\[ \wt D=\begin{pmatrix}1&0&0\\0&1&0\\0&0&1\end{pmatrix}. \]
This is the identity matrix. Following our criterion (\ref{criteria}), $\mbb Z_3$ of the $SU(3)_3$ WZW model is free of our anomaly.

\subsection{General WZWs}\label{SZZWZW}
In general $G_k$ WZW models, primaries are labeled by affine weights $\hat\mu$. When one puts the model on $\mbb T^2$ with modulus $\tau$, the twisted partition function $Z_{(h,h)}(\tau)$ is given by \cite{NY,CFT}
\begin{equation}
    Z_{(h,h)}(\tau)=\sum_{\hat\mu\in P^k_+}e^{-\pi ik|A\hat\omega_0|^2-2\pi i(A\hat\omega_0,\hat\mu)}\bar\chi_{A\hat\mu}(\bar\tau)\chi_{\hat\mu}(\tau),\label{Zhh}
\end{equation} where $A$ is the outer automorphism action corresponding to the element $h$ of the center of $G$ as mentioned before. 
Now the modular $S$-matrix elements satisfy
\begin{equation}
    S_{\hat\mu,A\hat\nu}=e^{-2\pi i(A\hat\omega_0,\hat\mu)}S_{\hat\mu,\hat\nu}.\label{prop}
\end{equation}
Using this, one can compute $SZ_{(h,h)}$;
\begin{equation}
\begin{split}
    SZ_{(h,h)}(\tau)&=S\sum_{\hat\mu\in P^k_+}e^{-\pi ik|A\hat\omega_0|^2-2\pi i(A\hat\omega_0,\hat\mu)}\bar\chi_{A\hat\mu}(\bar\tau)\chi_{\hat\mu}(\tau)\\
    &=\sum_{\hat\mu,\hat\nu_1,\hat\nu_2\in P^k_+}e^{-\pi ik|A\hat\omega_0|^2-2\pi i(A\hat\omega_0,\hat\mu)}S^*_{\hat\nu_1,A\hat\mu}\bar\chi_{\hat\nu_1}(\bar\tau)S_{\hat\nu_2,\hat\mu}\chi_{\hat\nu_2}(\tau)\\
    &=\sum_{\hat\mu,\hat\nu_1,\hat\nu_2\in P^k_+}e^{-\pi ik|A\hat\omega_0|^2+2\pi i(A\hat\omega_0,\hat\nu_1)}S^*_{\hat\nu_1,\hat\mu}\bar\chi_{\hat\nu_1}(\bar\tau)S_{A\hat\nu_2,\hat\mu}\chi_{\hat\nu_2}(\tau)\\
    &=\sum_{\hat\mu\in P^k_+}e^{-\pi ik|A\hat\omega_0|^2+2\pi i(A\hat\omega_0,A\hat\mu)}\bar\chi_{A\hat\mu}(\bar\tau)\chi_{\hat\mu}(\tau).
\end{split}\label{SZhh}
\end{equation}
The modular $S$-transformation exchanges the two cycles of the torus and one might naively expect that $S$ will preserve $Z_{(h,h)}$ because the twisting along two different cycles are the same. However, as we argued earlier, the possible ``ordering" of the insertions will obstruct the $S$-invariance of $Z_{(h,h)}$. This obstruction shows up as an anomaly. As proposed before, the examination of the $S$-transformation of $Z_{(h,h)}$ will tell us whether the symmetry is anomalous,
\[ SZ_{(h,h)}(\tau)\stackrel?=Z_{(h.h)}(\tau). \]
From (\ref{SZhh}) it is clear that generally there is a phase matrix mismatch after $S$-transformation. More concretely, since the (products of) characters $\bar\chi_{A\hat\mu}\chi_{\hat\mu}$ are the same, the difference between $SZ_{(h,h)}$ and $Z_{(h,h)}$ can only appear as a diagonal $|P^k_+|\times|P^k_+|$ phase matrix $D$ acting on $\bar\chi_{A\hat\mu}\chi_{\hat\mu}$'s, where $|P^k_+|$ is the total number of the WZW primaries. 
For our purpose to detect the anomaly we only concern the phase difference appearing as a $|\Gamma|\times|\Gamma|$ submatrix $\wt D$ of $D$ spanned by primary states $\hat\mu$ (and its $A$-transformations $A\hat\mu$) corresponding to the topological defect lines of $\Gamma$. 
Let us denote the subset of primary states which generate the center $\Gamma$ as $P^k_+\big|$, and the corresponding partition function as
\[ Z_{(h,h)}(\tau)\Big|:=\sum_{\hat\mu\in P^k_+\big|}e^{-\pi ik|A\hat\omega_0|^2-2\pi i(A\hat\omega_0,\hat\mu)}\bar\chi_{A\hat\mu}(\tau)\chi_{\hat\mu}(\tau). \]
There is no anomaly for $\Gamma$ iff
\begin{equation}
    SZ_{(h,h)}(\tau)\Big|=Z_{(h,h)}(\tau)\Big|.\label{anomfreecond}
\end{equation}
This is equivalent to (\ref{criteria}).
Therefore, to use our criterion, one has to evaluate the scalar product $(A\hat\omega_0,A\hat\mu)$, and compare the result with another $(A\hat\omega_0,\hat\mu)$. Using eq. (14.96) of \cite{CFT}, we get
\begin{align*}
    (A\hat\omega_0,A\hat\mu)&=\left(A\hat\omega_0,kA\hat\omega_0+\sum_{j=1}^r\mu_jA(\hat\omega_j-a_j^\vee\hat\omega_0)\right)\\
    &=\left(k-\sum_{j=1}^r\mu_ja_j^\vee\right)|A\hat\omega_0|^2+\sum_{j=1}^r\mu_j(A\hat\omega_0,A\hat\omega_j).
\end{align*}
We will use this formula repeatedly below. Notice that in the case of $\hat G=SU(2)$, since the rank is one, we have $\hat\mu=[\mu_0;\mu_1]$ and thus
\[ (A\hat\omega_0,\hat\mu)=\frac12\mu_1,\quad(A\hat\omega_0,A\hat\mu)=\frac12\mu_0=\frac12(k-\mu_1)=\frac12k-(A\hat\omega_0,\hat\mu)\ ,\]
which leads to
\[ e^{2\pi i(A\hat\omega_0,A\hat\mu)}=e^{\pi ik}e^{-2\pi i(A\hat\omega_0,\hat\mu)}\ . \]
Therefore
\[ SZ_{(h,h)}(\tau)=(-)^k\sum_{\hat\mu\in P^k_+}e^{-{k\pi i\over2}-2\pi i(A\hat\omega_0,\hat\mu)}\bar\chi_{A\hat\mu}(\tau)\chi_{\hat\mu}(\tau)=(-)^kZ_{(h,h)}(\tau)\ ,\]
which is exactly what we obtained previously for $SU(2)_k$.
In general cases, we can detect the anomaly by only computing the two scalar products,
\[ -2\pi i(A\hat\omega_0,\hat\mu) ~\text{and}~ +2\pi i(A\hat\omega_0,A\hat\mu)\]
or equivalently, by computing matrix elements~\footnote{Definitions of $D$- and $\wt D$-matrices are slightly different from the previous ones, however, they play the same roles in detecting our anomalies.} 
\begin{equation}\wt D_{\hat\mu\hat\mu}=e^{2\pi i(A\hat\omega_0,A\hat\mu+\hat\mu)}
\end{equation} in the truncated space.
We will adopt this method below.

\subsubsection{$A_r$ type i.e., $\mfrak{su}(r+1)$}
In this case, the center is a cyclic group $\Gamma=\mbb Z_{r+1}$. The fundamental element $A$ of the outer automorphism group $\Gamma$ acts as
\[ A[\mu_0;\mu_1,\cdots,\mu_{r-1},\mu_r]=[\mu_r;\mu_0,\cdots,\mu_{r-2},\mu_{r-1}]. \]
A small computation shows
\[ (A\hat\omega_0,\hat\mu)=-\frac1{r+1}\sum_{j=1}^rj\mu_j\quad\text{mod }1, \]
and
\[ (A\hat\omega_0,A\hat\mu)=-\frac k{r+1}-\frac1{r+1}\sum_{j=1}^rj\mu_j\quad\text{mod }1. \]
So we have
\[ (A\hat\omega_0,A\hat\mu)=-\frac k{r+1}+(A\hat\omega_0,\hat\mu)\quad\text{mod }1, \]
and we do not get $SZ_{(h,h)}=Z_{(h,h)}$ in general, meaning that there is an anomaly. To find the anomaly-free condition, we truncate the full $|P^k_+|\times|P^k_+|$ matrix $D$ to $|\Gamma|\times|\Gamma|=(r+1)\times(r+1)$ submatrix $\wt D$. Since the center $\Gamma=\mbb Z_{r+1}$ is generated by $\hat\mu=[k;0,\dots,0]$ and its cyclic permutations, we have
\begin{equation}
    \wt D=\text{diag}(e^{2\pi i\frac{-k}{r+1}},e^{2\pi i\frac{-3k}{r+1}},\dots,e^{2\pi i\frac{-k(2j+1)}{r+1}},\dots,e^{2\pi i\frac k{r+1}}).\label{A}
\end{equation}
This submatrix reduces to the identity matrix $1_{r+1}$ iff $k\in(r+1)\mbb Z$. This means part of the partition functions $Z_{(h,h)}$ and $SZ_{(h,h)}$ spanned by the generators of the center $\Gamma=\mbb Z_{r+1}$ are the same, and we interpret there is no anomaly. Thus the anomaly-free condition is given by $k\in(r+1)\mbb Z$.

\subsubsection{$B_r$ type i.e., $\mfrak{so}(2r+1)$}
In this case the center is $\Gamma=\mbb Z_2$, and the fundamental element $A$ of the group acts like
\[ A[\mu_0;\mu_1,\mu_2,\cdots,\mu_r]=[\mu_1;\mu_0,\mu_2,\cdots,\mu_r]. \]
One can compute
\[ (A\hat\omega_0,\hat\mu)=\sum_{j=1}^rF_{1j}\mu_j, \]
and
\[ (A\hat\omega_0,A\hat\mu)=k-\sum_{j=1}^rF_{1j}\mu_j=k-(A\hat\omega_0,\hat\mu), \]
where $F_{jk}$ is the quadratic form matrix. Therefore we have
\[ (A\hat\omega_0,A\hat\mu)=k-(A\hat\omega_0,\hat\mu)\quad\text{mod }1, \]
implying
\begin{equation}
    SZ_{(h,h)}(\tau)=e^{2\pi ik}Z_{(h,h)}(\tau),\label{B}
\end{equation}
and the partition function is invariant under the modular $S$-transformation iff $k\in\mbb Z$.

\subsubsection{$C_r$ type i.e., $\mfrak{sp}(2r)$}
Again the center is given by $\Gamma=\mbb Z_2$, and the fundamental element $A$ of the group maps
\[ A[\mu_0;\mu_1,\cdots,\mu_r]=[\mu_r;\mu_{r-1},\cdots,\mu_0]. \]
One obtains
\[ (A\hat\omega_0,\hat\mu)=\sum_{j=1}^rF_{rj}\mu_j, \]
and
\[ (A\hat\omega_0,A\hat\mu)=\frac{rk}2-\sum_{j=1}^rF_{rj}\mu_j=\frac{rk}2-(A\hat\omega_0,\hat\mu). \]
So we have
\[ (A\hat\omega_0,A\hat\mu)=\frac{rk}2-(A\hat\omega_0,\hat\mu), \]
implying
\begin{equation}
    SZ_{(h,h)}(\tau)=(-)^{rk}Z_{(h,h)}(\tau),\label{C}
\end{equation}
and the partition function is invariant under the modular $S$-transformation iff $rk\in2\mbb Z$.

\subsubsection{$D_r$ type i.e., $\mfrak{so}(2r)$}
The center groups are different depending on whether the rank $r$ is even or odd. We study them separately.

\underline{$r\in2\mbb Z$}~~ In this case, the center is given by $\Gamma=\mbb Z_2\times\mbb Z_2$. We denote nontrivial elements of each $\mbb Z_2$ by $A$ and $\wt A$, respectively. They acts like
\[ A[\mu_0;\mu_1,\mu_2,\cdots,\mu_{r-2},\mu_{r-1},\mu_r]=[\mu_1;\mu_0,\mu_2,\cdots,\mu_{r-2},\mu_r,\mu_{r-1}] \]
and
\[ \wt A[\mu_0;\mu_1,\mu_2,\cdots,\mu_{r-2},\mu_{r-1},\mu_r]=[\mu_r;\mu_{r-1},\mu_{r-2},\cdots,\mu_2,\mu_1,\mu_0]. \]
There are three $\mbb Z_2$ subgroups of the center corresponding to $A,\wt A$, and $\wt AA$. Their scalar products are given by
\begin{align*}
	(A\hat\omega_0,\hat\mu)&=\sum_{j=1}^rF_{1j}\mu_j,\\
	(A\hat\omega_0,A\hat\mu)&=k-\sum_{j=1}^rF_{1j}\mu_j,\\
	(\wt A\hat\omega_0,\hat\mu)&=\sum_{j=1}^rF_{rj}\mu_j,\\
	(\wt A\hat\omega_0,\wt A\hat\mu)&=\frac{rk}4-\sum_{j=1}^rF_{rj}\mu_j,\\
	(\wt AA\hat\omega_0,\hat\mu)&=\sum_{j=1}^{r-2}\frac j2\mu_j+\frac r4\mu_{r-1}+\frac{r-2}4\mu_r,\\
	(\wt AA\hat\omega_0,\wt AA\hat\mu)&=\frac{rk}4-\sum_{j=1}^{r-2}\frac j2\mu_j-\frac r4\mu_{r-1}-\frac{r-2}4\mu_r.
\end{align*}
So we have
\[ (A\hat\omega_0,A\hat\mu)=k-(A\hat\omega_0,\hat\mu),\quad(\wt A\hat\omega_0,\wt A\hat\mu)=\frac{rk}4-(\wt A\hat\omega_0,\hat\mu),\quad(\wt AA\hat\omega_0,\wt AA\hat\mu)=\frac{rk}4-(\wt AA\hat\omega_0,\hat\mu). \]
Therefore the twisted partition functions obey
\begin{align}
    SZ_{(h,h)}(\tau)&=e^{2\pi ik}Z_{(h,h)}(\tau).\label{DevenA}\\
    SZ_{(\wt h,\wt h)}(\tau)&=(-)^{lk}Z_{(\wt h,\wt h)}(\tau),\label{DeventildeA}\\
    SZ_{(\wt hh,\wt hh)}(\tau)&=(-)^{lk}Z_{(\wt hh,\wt hh)}(\tau).\label{DeventildeAA}
\end{align}
where $r=2l$. To achieve the full invariance of the partition function, we have to impose all phases in (\ref{DevenA}), (\ref{DeventildeA}) , and (\ref{DeventildeAA}) be trivial, resulting in $k\in2\mbb Z$ if $r\in4\mbb Z+2$, and $k\in\mbb Z$ if $r\in4\mbb Z$. In other words, $lk\in2\mbb Z$.

\underline{$r\in2\mbb Z+1$}~~ The center is given by $\Gamma=\mbb Z_4$. The fundamental element $A$ of the group maps
\[ A[\mu_0;\mu_1,\mu_2,\cdots,\mu_{r-2},\mu_{r-1},\mu_r]=[\mu_{r-1};\mu_r,\mu_{r-2},\cdots,\mu_2,\mu_1,\mu_0]. \]
Since the group is larger than $\mbb Z_2$, we have to truncate the matrix $D$. $\wt D$ is spanned by generators of $\mbb Z_4$, i.e., $\hat\mu=[k;0,\dots,0]$ and its $A$-transformations. A straightforward computation gives
\[ (A\hat\omega_0,\hat\mu)=\sum_{j=1}^r\mu_jF_{r-1,j}=\sum_{j=1}^{r-2}\frac j2\mu_j+\frac r4\mu_{r-1}+\frac{r-2}4\mu_r, \]
and
\[ (A\hat\omega_0,A\hat\mu)=\frac{rk}4-\sum_{j=1}^{r-2}\frac j2\mu_j-\frac{r-2}4\mu_{r-1}-\frac r4\mu_r, \]
where $r=2l+1$.
So we have
\[ (A\hat\omega_0,A\hat\mu)=\frac{rk}4+(A\hat\omega_0,\hat\mu)-l\mu_{r-1}-l\mu_r\quad\text{mod }1=\frac{rk}4+(A\hat\omega_0,\hat\mu)\quad\text{mod }1, \]
resulting in the submatrix $\wt D$
\begin{equation}
    \wt D=\text{diag}((-)^{\frac k2(2l+1)},(-)^{k+\frac k2(2l-1)},(-)^{\frac k2(2l-1)},(-)^{k+\frac k2(2l+1)}),\label{Dodd}
\end{equation}
where $r=2l+1$ with $l\in\mbb Z$. Therefore, iff $k\in4\mbb Z$, there is no anomaly.

\subsubsection{$E_6$}
The center is given by $\Gamma=\mbb Z_3$. The fundamental element $A$ of the group acts like
\[ A[\mu_0;\mu_1,\mu_2,\mu_3,\mu_4,\mu_5,\mu_6]=[\mu_5;\mu_0,\mu_6,\mu_3,\mu_2,\mu_1,\mu_4]. \]
A small computation shows
\[ (A\hat\omega_0,\hat\mu)=\frac43\mu_1+\frac53\mu_2+\frac63\mu_3+\frac43\mu_4+\frac23\mu_5+\frac33\mu_6, \]
while
\begin{align*}
    (A\hat\omega_0,A\hat\mu)&=\frac43(k-\mu_1-2\mu_2-3\mu_3-2\mu_4-\mu_5-2\mu_6)+\frac23\mu_1+\frac43\mu_2+\frac63\mu_3+\frac33\mu_4+\frac53\mu_6\\
    &=\frac43k-\frac23\mu_1-\frac43\mu_2-\frac63\mu_3-\frac53\mu_4-\frac43\mu_5-\frac33\mu_6\\
    &=\frac43k+(A\hat\omega_0,\hat\mu)-2\mu_1-3\mu_2-4\mu_3-3\mu_4-2\mu_5-2\mu_6\\
    &=\frac43k+(A\hat\omega_0,\hat\mu)\quad\text{mod }1.
\end{align*}
Since the center is larger than $\mbb Z_2$, we have to truncate $D$ to $\wt D$ spanned by its generators, i.e., $\hat\mu=[k;0,0,0,0,0,0]$ and its $A$-transformations. For $A$ such that $A\hat\omega_0=\hat\omega_1$, the submatrix is then given by
\begin{equation}
    \wt D=\text{diag}(e^{2\pi ik/3},1,e^{2\pi i\cdot2k/3}).\label{E6}
\end{equation}
This matrix reduces to $1_3$ iff $k\in3\mbb Z$, giving the anomaly-free condition.

\subsubsection{$E_7$}
The center is given by $\Gamma=\mbb Z_2$, The fundamental element $A$ of the group maps
\[ A[\mu_0;\mu_1,\mu_2,\mu_3,\mu_4,\mu_5,\mu_6,\mu_7]=[\mu_6;\mu_5,\mu_4,\mu_3,\mu_2,\mu_1,\mu_0,\mu_7]. \]
One can easily show
\[ (A\hat\omega_0,\hat\mu)=\mu_1+2\mu_2+3\mu_3+\frac52\mu_4+2\mu_5+\frac32\mu_6+\frac32\mu_7, \]
and
\[ (A\hat\omega_0,A\hat\mu)=\frac32\mu_0+\sum_{j=1}^7\mu_j(\hat\omega_6,A\hat\omega_j). \]
So we have
\[ (A\hat\omega_0,A\hat\mu)=\frac32k-(A\hat\omega_0,\hat\mu), \]
implying
\begin{equation}
    SZ_{(h,h)}(\tau)=(-)^{3k}Z_{(h,h)}(\tau),\label{E7}
\end{equation}
and the partition function is invariant under the modular $S$-transformation iff $k\in2\mbb Z$.

\subsection{Minimal models}
Our consideration so far was restricted to WZW models, which are special class of diagonal RCFTs. A explicit family of TDLs in diagonal RCFTs are called Verlinde lines \cite{Verlinde:1988sn}. Due to the modular invariance, there is a one-to-one correspondence between Verlinde lines and chiral vertex algebra primaries \cite{Moore:1988qv,Moore:1989yh,Moore:1989vd}. The TDLs associated to global symmetry are called invertible lines. One can repeat our previous computation using these invertible lines, namely twist the torus partition function by invertible lines and consider their modular $S$-transformations to detect our anomalies. It is therefore natural to expect that our criterion works for general diagonal RCFTs. In this section, we test our proposal by studying some minimal models.

\subsubsection{Critical Ising model, i.e., $M(4,3)$}
The first canonical example is the critical Ising model. It has three primary operators $id,\varepsilon,$ and $\sigma$. The theory has $\mbb Z_2$ symmetry, generated by the topological line associated to $\varepsilon$. The twisted partition function is given by
\[ Z_{(h,1)}(\tau)=|\chi_{id}(\tau)|^2+|\chi_\varepsilon(\tau)|^2-|\chi_\sigma(\tau)|^2 \]
because just $\sigma$ is odd under the $\mbb Z_2$. Performing the modular $S$-transformation, one obtains
\[ Z_{(1,h)}(\tau)=\bar\chi_{id}(\bar\tau)\chi_\varepsilon(\tau)+\bar\chi_\varepsilon(\bar\tau)\chi_{id}(\tau)+|\chi_\sigma(\tau)|^2. \]
A quicker way to compute the twisted partition function $Z_{(1,h)}$ is to use the fusion coefficients ${N_{ki}}^j$ (or equivalently fusion rules). The partition function can be written as a trace over the twisted Hilbert space $\mcal H_{\mcal L}$ which is defined by inserting the corresponding topological line $\mcal L$ along the time direction:
\[ Z_{(1,h)}(\tau)=\tr_{\mcal H_{\mcal L}}\Big(q^{L_0-c/24}\bar q^{\bar L_0-c/24}\Big) \]
where $L_0$ and $\bar L_0$ are the usual Virasoro generators, and $q:=e^{2\pi i\tau}$. Then the partition function can be calculated with ease because it is given by
\[ Z_{(1,h_k)}(\tau)=\sum_{i,j}{N_{ki}}^j\chi_i(\tau)\bar\chi_j(\bar\tau) \]
where $h_k$ is the group element generated by the topological line $\mcal L_k$ associated to the primary operator $\phi_k$. Then using the fusion rules
\[ \varepsilon\times\varepsilon=id,\quad\varepsilon\times id=\varepsilon,\quad\varepsilon\times\sigma=\sigma, \]
one can easily reproduce the twisted partition function $Z_{(1,h)}$ above. We will use this quicker way below.

Finally, by performing the modular $T$-transformation once, we get
\begin{equation}
    Z_{(h,h)}(\tau)=-\bar\chi_{id}(\bar\tau)\chi_\varepsilon(\tau)-\bar\chi_\varepsilon(\bar\tau)\chi_{id}(\tau)+\bar\chi_\sigma(\bar\tau)\chi_\sigma(\tau).\label{Ising}
\end{equation}
The reduced modular $S$-matrix method~\cite{HWZ} claims the $\mbb Z_2$ is free of an anomaly. So one would expect the twisted partition function is invariant under the modular $S$-transformation, and indeed one can show
\begin{equation}
    SZ_{(h,h)}(\tau)=Z_{(h,h)}(\tau).\label{IsingSZZ}
\end{equation}

\subsubsection{Tricritical Ising model, i.e., $M(5,4)$}
In the same way, one can study the tricritical Ising model. The theory has six primary operators $id,\sigma',\sigma,\varepsilon'',\varepsilon',$ and $\varepsilon$. This theory also has a $\mbb Z_2$ symmetry which is generated by the topological line associated to $\varepsilon''$. Using the fusion rules involving $\varepsilon''$, the twisted partition function can be calculated easily:
\[ Z_{(1,h)}(\tau)=|\chi_{\sigma'}(\tau)|^2+|\chi_\sigma(\tau)|^2+\bar\chi_{id}(\bar\tau)\chi_{\varepsilon''}(\tau)+\bar\chi_{\varepsilon''}(\bar\tau)\chi_{id}(\tau)+\bar\chi_{\varepsilon'}(\bar\tau)\chi_\varepsilon(\tau)+\bar\chi_\varepsilon(\bar\tau)\chi_{\varepsilon'}(\tau). \]
Then performing the modular $T$-transformation once, one obtains
\begin{equation}
    Z_{(h,h)}(\tau)=|\chi_{\sigma'}(\tau)|^2+|\chi_\sigma(\tau)|^2-\bar\chi_{id}(\bar\tau)\chi_{\varepsilon''}(\tau)-\bar\chi_{\varepsilon''}(\bar\tau)\chi_{id}(\tau)-\bar\chi_{\varepsilon'}(\bar\tau)\chi_\varepsilon(\tau)-\bar\chi_\varepsilon(\bar\tau)\chi_{\varepsilon'}(\tau).\label{tricriticalIsing}
\end{equation}
The twisted partition function is invariant under the modular $S$-transformation
\begin{equation}
    SZ_{(h,h)}(\tau)=Z_{(h,h)}(\tau)\label{tricriticalIsingSZZ}
\end{equation}
consistent with the reduced modular $S$-matrix method.

\subsubsection{Tetracritical Ising model, i.e., $M(6,5)$}
This model has 10 primary operators $\{1,u,f,v,w,\hat w,\hat v,\hat f,\hat u,\hat1\}$ following the convention of \cite{FFRS}. Among these, $w$ with
\[ w\times w=1 \]
corresponds to an invertible Verlinde line $C$, which generates the $\mbb Z_2$ of the model. To explore whether the $\mbb Z_2$ has an anomaly, we consider partition functions twisted by $C$. The first twisted partition function $Z_{(1,h)}$ is given by
\[ Z_{(1,h)}(\tau)=\bar\chi_1\chi_w+\bar\chi_w\chi_1+\bar\chi_u\chi_v+\bar\chi_v\chi_u+|\chi_f|^2+\bar\chi_{\hat w}\chi_{\hat1}+\bar\chi_{\hat1}\chi_{\hat w}+\bar\chi_{\hat v}\chi_{\hat u}+\bar\chi_{\hat u}\chi_{\hat v}+|\chi_{\hat f}|^2. \]
Performing the modular $T$-transformation, one obtains $Z_{(h,h)}$:
\[ Z_{(h,h)}(\tau)=\bar\chi_1\chi_w+\bar\chi_w\chi_1-\bar\chi_u\chi_v-\bar\chi_v\chi_u+|\chi_f|^2+\bar\chi_{\hat w}\chi_{\hat1}+\bar\chi_{\hat1}\chi_{\hat w}-\bar\chi_{\hat v}\chi_{\hat u}-\bar\chi_{\hat u}\chi_{\hat v}+|\chi_{\hat f}|^2. \]
One can check this twisted partition function is invariant under the modular $S$-transformation
\[ SZ_{(h,h)}(\tau)=Z_{(h,h)}(\tau), \]
implying the $\mbb Z_2$ is free of our anomaly, consistent with the reduced modular $S$-matrix method.

\subsubsection{Three-state Potts model, i.e., $M(6,5)$}
Viewed as representations of the Virasoro algebra, the theory has 12 primary operators $id,\varepsilon,X,Y,\Phi,\wt\Phi,\Omega,\wt\Omega,\sigma_1,\sigma_2,Z_1,$ and $Z_2$. Among them, $\Omega$ and $\wt\Omega$ have integer scaling dimensions, implying larger symmetry. In fact, it is known that the theory has an extended symmetry $W_3$. Viewed as representations of the $W_3$ algebra, the theory is block diagonal, and we can use the power of Verlinde lines. To study anomalies of the theory, it is convenient to view the theory as a diagonal RCFT. Then characters of the theory are combined into six blocks
\begin{equation}
\begin{split}
    \chi_{C_{11}}:=\chi_{id}+\chi_Y,&\quad\chi_{C_{21}}:=\chi_\varepsilon+\chi_X,\\
    \chi_{C_{13}^{(1)}}=\chi_{Z_1},\quad\chi_{C_{13}^{(2)}}=\chi_{Z_2},&\quad\chi_{C_{23}^{(1)}}=\chi_{\sigma_1},\quad\chi_{C_{23}^{(2)}}=\chi_{\sigma_2}.
\end{split}\label{extcharacter}
\end{equation}

The theory has $S_3$ symmetry, whose subgroups are $\mbb Z_3$ generated by the topological line associated with $Z_1$ or $C_{13}^{(1)}$, and the charge conjugation $\mbb Z_2$. %generated by the topological line associated with $Y$.
We would like to study whether the $\mbb Z_3\subset S_3$ has an anomaly. Using the extended fusion rules, one obtains
\begin{align*}
    Z_{(1,h)}(\tau)&=\bar\chi_{C_{11}}(\bar\tau)\chi_{C_{13}^{(2)}}(\tau)+\bar\chi_{C_{21}}(\bar\tau)\chi_{C_{23}^{(2)}}(\tau)+\bar\chi_{C_{13}^{(1)}}(\bar\tau)\chi_{C_{11}}(\tau)\\
    &~~~+\bar\chi_{C_{13}^{(2)}}(\bar\tau)\chi_{C_{13}^{(1)}}(\tau)+\bar\chi_{C_{23}^{(1)}}(\bar\tau)\chi_{C_{21}}(\tau)+\bar\chi_{C_{23}^{(2)}}(\bar\tau)\chi_{C_{23}^{(1)}}(\tau).
\end{align*}
Performing the modular $T$-transformation once, one obtains
\begin{equation}
\begin{split}
    Z_{(h,h)}(\tau)&=\omega^2\bar\chi_{C_{11}}(\bar\tau)\chi_{C_{13}^{(2)}}(\tau)+\omega^2\bar\chi_{C_{21}}(\bar\tau)\chi_{C_{23}^{(2)}}(\tau)+\omega\bar\chi_{C_{13}^{(1)}}(\bar\tau)\chi_{C_{11}}(\tau)\\
    &~~~+\bar\chi_{C_{13}^{(2)}}(\bar\tau)\chi_{C_{13}^{(1)}}(\tau)+\omega\bar\chi_{C_{23}^{(1)}}(\bar\tau)\chi_{C_{21}}(\tau)+\bar\chi_{C_{23}^{(2)}}(\bar\tau)\chi_{C_{23}^{(1)}}(\tau),
\end{split}\label{3statePotts}
\end{equation}
where $\omega=e^{2\pi i/3}$. The modular $S$-matrix acting on the extended characters is given by \cite{CFT}
\begin{equation}
    S=\frac{2\sin(\pi/5)}{\sqrt{15}}\begin{pmatrix}1&\zeta&1&1&\zeta&\zeta\\\zeta&-1&\zeta&\zeta&-1&-1\\1&\zeta&\omega&\omega^2&\omega\zeta&\omega^2\zeta\\1&\zeta&\omega^2&\omega&\omega^2\zeta&\omega\zeta\\\zeta&-1&\omega\zeta&\omega^2\zeta&-\omega&-\omega^2\\\zeta&-1&\omega^2\zeta&\omega\zeta&-\omega^2&-\omega\end{pmatrix}\label{PottsS}
\end{equation}
in the basis $\{C_{11},C_{21},C_{13}^{(1)},C_{13}^{(2)},C_{23}^{(1)},C_{23}^{(2)}\}$, where $\zeta:=\frac{1+\sqrt5}2$. Using the modular $S$-matrix, one can compute $SZ_{(h,h)}$ to find
\begin{equation}
\begin{split}
    SZ_{(h,h)}(\tau)&=\omega\bar\chi_{C_{11}}(\bar\tau)\chi_{C_{13}^{(1)}}(\tau)+\omega\bar\chi_{C_{21}}(\bar\tau)\chi_{C_{23}^{(1)}}(\tau)+\bar\chi_{C_{13}^{(1)}}(\bar\tau)\chi_{C_{13}^{(2)}}(\tau)\\
    &~~~+\omega^2\bar\chi_{C_{13}^{(2)}}(\bar\tau)\chi_{C_{11}}(\tau)+\bar\chi_{C_{23}^{(1)}}(\bar\tau)\chi_{C_{23}^{(2)}}(\tau)+\omega^2\bar\chi_{C_{23}^{(2)}}(\bar\tau)\chi_{C_{21}}(\tau).
\end{split}\label{3statePottsSZ}
\end{equation}
Since $\mbb Z_3$ is larger than $\mbb Z_2$, our criterion claims we should truncate the partition functions to contributions coming just from generators of the $\mbb Z_3$. Then one finds
\[ Z_{(h,h)}(\tau)\Big|=\omega^2\bar\chi_{C_{11}}(\bar\tau)\chi_{C_{13}^{(2)}}(\tau)+\omega\bar\chi_{C_{13}^{(1)}}(\bar\tau)\chi_{C_{11}}(\tau)+\bar\chi_{C_{13}^{(2)}}(\bar\tau)\chi_{C_{13}^{(1)}}(\tau) \]
or\footnote{As is clear from the construction, whether $\wt D$ is the identity or not, i.e., whether there exists an anomaly or not, does not depend on whether we use the full unitary matrix to define the $D$-matrix and then truncate, or we truncate the partition function first and read off unitary matrices to construct the $\wt D$-matrix directly. In this example, one can explicitly see the two methods give the same $\wt D$-matrix (\ref{3stateD}).}
\[ Z_{(h,h)}\Big|\lr\begin{pmatrix}0&\omega&0\\0&0&1\\\omega^2&0&0\end{pmatrix}=:U \]
in the basis $\{C_{11},C_{13}^{(1)},C_{13}^{(2)}\}$ and
\[ SZ_{(h,h)}(\tau)\Big|=\omega\bar\chi_{C_{11}}(\bar\tau)\chi_{C_{13}^{(1)}}(\tau)+\bar\chi_{C_{13}^{(1)}}(\bar\tau)\chi_{C_{13}^{(2)}}(\tau)+\omega^2\bar\chi_{C_{13}^{(2)}}(\bar\tau)\chi_{C_{11}}(\tau), \]
or
\[ SZ_{(h,h)}\Big|\lr\begin{pmatrix}0&0&\omega^2\\\omega&0&0\\0&1&0\end{pmatrix}=:U'. \]
Since the $\wt D$-matrix is given by
\begin{equation}
    \wt D:=U^{-1} U'=\begin{pmatrix}0&\omega&0\\0&0&\omega\\\omega&0&0\end{pmatrix},\label{3stateD}
\end{equation}
our criterion (\ref{criteria}) states that the $\mbb Z_3$ global symmetry of the three-state Potts model has an anomaly. This result is consistent with the reduced modular $S$-matrix method~\cite{HWZ}. By reducing the extended modular $S$-matrix (\ref{PottsS}) to the submatrix $\hat S$ spanned by $\{C_{11},C_{13}^{(1)},C_{13}^{(2)}\}$, one obtains the density matrix $\rho$
\[ \rho={\hat S\hat S^\dagger\over \text{Tr} \hat S\hat S^\dagger} =\frac13\begin{pmatrix}1&0&0\\0&1&0\\0&0&1\end{pmatrix}, \]
which has von Neumann entropy
\[ -\tr\left(\rho\ln\rho\right)=\ln3, \]
indicating that the $\mbb Z_3$ has an anomaly beyond WZW models.

%Let us come back to the relation between 't Hooft anomaly-free condition and the orbifoldability (\ref{anomorb}). If the two conditions were equivalent as usually stated in the literature, the three-state Potts model could not have well-defined $\mbb Z_3$ orbifold parition function due to the 't Hooft anomaly in the symmetry as we have just seen. A straightforward computation shows the orbifold partition function is given by
\begin{equation}
    \wt Z^{\mbb Z_3}=|\chi_{C_{11}}|^2+|\chi_{C_{21}}|^2+\bar\chi_{C_{13}^{(1)}}\chi_{C_{13}^{(2)}}+\bar\chi_{C_{13}^{(2)}}\chi_{C_{13}^{(1)}}+\bar\chi_{C_{23}^{(1)}}\chi_{C_{23}^{(2)}}+\bar\chi_{C_{23}^{(2)}}\chi_{C_{23}^{(1)}}.\label{3statePottsZ3orb}
\end{equation}
%which is modular invariant. Therefore, this model proves an existence of $H$-orbifoldable theory even when $H$ has a 't Hooft anomaly. Thus we believe (\ref{anomorb}) is true for diagonal RCFTs not just for WZW models but also for more general RCFTs such as minimal models.

\subsection{Interpretation of our anomaly}
So far, we have just stated that our criterion (\ref{criteria0}) detects some anomaly without interpreting what this anomaly is. We propose the anomaly is a mixed anomaly\footnote{We thank Yuji Tachikawa for pointing our erroneous interpretation in v1.} between the internal discrete symmetry and its ``$S$-dual." For WZW theories, the former is nothing but the center, and the latter is the outer automorphism. In the case of minimal models, only the former is familiar, but we can find the analogy of ``outer automorphism''.
Let us start from WZW models. We consider $SU(3)_k$ WZW models. Orbifold partition functions of the $SU(3)_1$ WZW model twisted by $\mbb Z_3$ is given by
\begin{equation}
\wt Z^{\mbb Z_3}=|\chi_1|^2+\bar\chi_{\bar3}\chi_3+\bar\chi_3\chi_{\bar3}.\label{su3_1orb}
\end{equation}
One can see this is not invariant under cyclic permutations $1\to3\to\bar3\to1$, the outer automorphisms. Since it is believed that orbifolding and gauging are the same, the theory (\ref{su3_1orb}) is obtained by gauging the $\mbb Z_3$ center. Gauging one symmetry and another is broken signals a mixed anomaly between the two. So this example is consistent with our proposal. Next, let us study the $SU(3)_2$ WZW model. The orbifold partition function is given by
\begin{equation}
\wt Z^{\mbb Z_3}=|\chi_{[2;0,0]}|^2+\bar\chi_{[1;0,1]}\chi_{[1;1,0]}+\bar\chi_{[1;1,0]}\chi_{[1;0,1]}+\bar\chi_{[0;0,2]}\chi_{[0;2,0]}+|\chi_{[0;1,1]}|^2+\bar\chi_{[0;2,0]}\chi_{[0;0,2]}.\label{su3_2orb}
\end{equation}
Again the outer automorphism is broken, consistent with our interpretation. Finally, let us examine the $SU(3)_3$ WZW model. The orbifold partition function is computed as
\begin{equation}
\wt Z^{\mbb Z_3}=|\chi_{[3;0,0]}+\chi_{[0;3,0]}+\chi_{[0;0,3]}|^2+3|\chi_{[1;1,1]}|^2.\label{su3_3orb}
\end{equation}
In this case, one can see the outer automorphism is preserved, supporting our claim.

What  about minimal models? Although the ``$S$-duals" of discrete internal global symmetries in these models are not known to our best knowledge, we can find them. In WZW models, one can read outer automorphisms from the fusion rules among the primaries. Following the same step, we can find ``$S$-duals" of the symmetries in minimal models. For example, the critical Ising model has $\mbb Z_2$ generated by $\mcal L_\varepsilon$. Looking at the fusion rules involving $\varepsilon$, we observe it effectively exchanges $id$ and $\varepsilon$. This is the automorphism we are interested. Since we have seen the model is free of our anomaly, the orbifold partition function should be invariant under the exchange. In deed, one can see
\begin{equation}
\wt Z^{\mbb Z_2}=|\chi_{id}|^2+|\chi_\varepsilon|^2+|\chi_\sigma|^2=Z_{(1,1)}\label{criticalIsingorb}
\end{equation}
is invariant under the exchange, consistent with our proposal. Similarly in the three-state Potts model, one finds that multiplication by $C_{13}^{(1)}$ effectively causes cyclic permutations $C_{11}\to C_{13}^{(1)}\to C_{13}^{(2)}\to C_{11}$ and $C_{21}\to C_{23}^{(1)}\to C_{23}^{(2)}\to C_{21}$, which is broken in (\ref{3statePottsZ3orb}), again consistent with our claim. We have also checked our proposal holds in tricritical and tetracritical Ising models.

\section{Invariant boundary states}\label{IB}
As we have explained in the introduction, anomalies and boundaries are closely related. To make the relation more precise, we turn to the boundary states of CFTs in this section. It is known that there is a physical basis called Cardy states \cite{C89}. As realized in \cite{HTHR,NY}, an existence of symmetry invariant Cardy states give anomaly-free conditions. Let us first review how the Cardy states are defined.

As we reviewed in the beginning of the section \ref{examples}, conformal families of WZW models are labeled by affine weights $\hat\mu\in P^k_+$. They provide primary states $|\hat\mu,\hat\mu\ra$. Linear combinations of the primary states and their descendants define what is called Ishibashi states \cite{Ishibashistate}
\[ |\hat\mu\ra\hspace{-2pt}\ra. \]
The Cardy states are constructed out of the Ishibashi states as \cite{C89}
\begin{equation}
    |\hat\mu\ra_c:=\sum_{\hat\lambda\in P^k_+}\frac{S_{\hat\mu\hat\lambda}}{\sqrt{S_{\hat0\hat\lambda}}}|\hat\lambda\ra\hspace{-2pt}\ra.\label{Cardystate}
\end{equation}
Under a center element $h\in B(G)$, it is mapped to
\[ h:|\hat\mu\ra_c\mapsto|A\hat\mu\ra_c, \]
where $A$ is the corresponding element of the outer automorphism group. Therefore, if there exists an invariant Cardy state, the affine Dynkin labels cannot be arbitrary, and the constraint can be translated to a condition on the level $k$. More concretely, if an affine weight $\hat\mu$ provides an invariant Cardy state under $h$, it must obey $|A\hat\mu\ra_c=|\hat\mu\ra_c$, or equivalently
\[ A\hat\mu=\hat\mu. \]
With the explicit action of $A$, this condition constrains affine Dynkin labels. We will illustrate the constraints for each algebra. The following computation essentially follows \cite{NY} where they used $Ch$, the charge conjugation $C$ times center symmetry. Here we basically repeat the same computation but with $h$. In the end, we focus on the relation between anomalies and the existence of invariant boundary states to support the recent conjecture \cite{Y19} that when there exists an $G$-invariant boundary state, $G$ is anomaly decoupled.

\subsection{$A_r$ type i.e., $\mfrak{su}(r+1)$}
Since $\Gamma=\mbb Z_{r+1}$, $A$ rotates Dynkin labels cyclically
\[ A[\mu_0;\mu_1,\cdots,\mu_{r-1},\mu_r]=[\mu_r;\mu_0,\cdots,\mu_{r-2},\mu_{r-1}]. \]
Equating this with the original Dynkin label, we obtain the constraints
\[ \mu_0=\mu_1=\cdots=\mu_r. \]
Thus the level is given by
\begin{equation}
    k\equiv\mu_0+\sum_{j=1}^r\mu_j=(r+1)\mu_0\in(r+1)\mbb Z,\label{Ak}
\end{equation}
reproducing the anomaly-free condition $k\in(r+1)\mbb Z$. This result is true even if $h$ is raised to any power $s$ such that $\text{gcd}(s,r+1)=1$, i.e., if $h^s$ still generates $\mbb Z_{r+1}$.

\subsection{$B_r$ type i.e., $\mfrak{so}(2r+1)$}
$A$ acts as
\[ A[\mu_0;\mu_1,\mu_2,\cdots,\mu_r]=[\mu_1;\mu_0,\mu_2,\cdots,\mu_r]. \]
Thus $A\hat\mu=\hat\mu$ requires
\[ \mu_0=\mu_1. \]
So we have the level
\begin{equation}
    k\equiv\mu_0+\mu_1+2\sum_{j=2}^{r-1}\mu_j+\mu_r=2\left(\mu_0+\sum_{j=2}^{r-1}\mu_j\right)+\mu_r\in\mbb Z,\label{Bk}
\end{equation}
reproducing the anomaly-free condition $k\in\mbb Z$.

\subsection{$C_r$ type i.e., $\mfrak{sp}(2r)$}
$A$ maps
\[ A[\mu_0;\mu_1,\cdots,\mu_r]=[\mu_r;\mu_{r-1},\cdots,\mu_0]. \]
Thus $A\hat\mu=\hat\mu$ imposes the following conditions simultaneously:
\[ \mu_0=\mu_r\&\mu_1=\mu_{r-1}\&\cdots. \]
This condition depends on whether $r$ is odd or even.

\underline{$r\in2\mbb Z+1$}~~ If $r$ is odd, there are even numbers of components, resulting in
\[ \mu_0=\mu_r\&\mu_1=\mu_{r-1}\&\cdots\&\mu_l=\mu_{l+1}, \]
where $r=2l+1$. Therefore the level is given by
\begin{equation}
    k\equiv\mu_0+\sum_{j=1}^r\mu_j=2(\mu_0+\mu_1+\cdots+\mu_l)\in2\mbb Z.\label{Codd}
\end{equation}

\underline{$r\in2\mbb Z$}~~ If $r$ is even, there are odd numbers of components, and the one in the middle is free:
\[ \mu_0=\mu_r\&\mu_1=\mu_{r-1}\&\cdots\&\mu_{l-1}=\mu_{l+1}\&\mu_l\text{ is free}, \]
where $r=2l$. Thus the level is given by
\begin{equation}
    k\equiv\mu_0+\sum_{j=1}^r\mu_j=2(\mu_0+\mu_1+\cdots+\mu_{l-1})+\mu_l\in\mbb Z.\label{Ceven}
\end{equation}
Combining the results (\ref{Codd}) and (\ref{Ceven}), we reproduce the anomaly-free condition $rk\in2\mbb Z$.

\subsection{$D_r$ type i.e., $\mfrak{so}(2r)$}
\subsubsection{$r\in2\mbb Z$}
In this case, there are two nontrivial elements of the center $h$ and $\wt h$, and correspondingly there are two nontrivial elements of the outer automorphism group $A$ and $\wt A$, respectively. Let us first consider $A$. It maps
\[ A[\mu_0;\mu_1,\mu_2,\cdots,\mu_{r-2},\mu_{r-1},\mu_r]=[\mu_1;\mu_0,\mu_2,\cdots,\mu_{r-2},\mu_r,\mu_{r-1}]. \]
Thus $A\hat\mu=\hat\mu$ requires
\[ \mu_0=\mu_1\&\mu_{r-1}=\mu_r. \]
%resulting in
%\begin{equation}
%    k\equiv\mu_0+\mu_1+2\sum_{j=2}^{r-2}\mu_j+\mu_{r-1}+\mu_r=2\left(\mu_0+\sum_{j=2}^{r-2}\mu_j+\mu_{r-1}\right)\in2\mbb Z.\label{DevenAk}
%\end{equation}
Next, let us consider $\wt A$. This maps
\[ \wt A[\mu_0;\mu_1,\mu_2,\cdots,\mu_{r-2},\mu_{r-1},\mu_r]=[\mu_r;\mu_{r-1},\mu_{r-2},\cdots,\mu_2,\mu_1,\mu_0]. \]
$\wt A\hat\mu=\hat\mu$ requires
\[ \mu_0=\mu_r\&\mu_1=\mu_{r-1}\&\cdots\&\mu_{l-1}=\mu_{l+1}\&\mu_l\text{ is free}, \]
where $r=2l$.
%Thus the level is given by
%\begin{equation}
%    k\equiv\mu_0+\mu_1+2\sum_{j=2}^{r-2}\mu_j+\mu_{r-1}+\mu_r=2\left(\mu_0+\mu_1+2\sum_{j=2}^{l-1}\mu_j+\mu_l\right)\in2\mbb Z.\label{DeventildeAk}
%\end{equation}
%Combining the two conditions, we reproduce the anomaly-free condition $k\in2\mbb Z$.
To have a boundary state invariant under both $A$ and $\wt A$, the affine weight must thus obey
\[ \mu_0=\mu_1=\mu_{r-1}=\mu_r\&\mu_2=\mu_{r-2}\&\mu_3=\mu_{r-3}\&\cdots\&\mu_{l-1}=\mu_{l+1}\&\mu_l\text{ is free}, \]
resulting in
\begin{equation}
    k\equiv\mu_0+\mu_1+2\sum_{j=2}^{r-2}\mu_j+\mu_{r-1}+\mu_r=4\left(\mu_0+\sum_{j=2}^{l-1}\mu_j\right)+2\mu_l\in2\mbb Z.\label{Devenk}
\end{equation}
Thus there exists a boundary state which is invariant under whole of the center $\mbb Z_2\times\mbb Z_2$ iff $k\in2\mbb Z$.

\subsubsection{$r\in2\mbb Z+1$}
In this case, $A$ maps
\[ A[\mu_0;\mu_1,\mu_2,\cdots,\mu_{r-2},\mu_{r-1},\mu_r]=[\mu_{r-1};\mu_r,\mu_{r-2},\cdots,\mu_2,\mu_1,\mu_0]. \]
$A\hat\mu=\hat\mu$ requires
\[ \mu_0=\mu_1=\mu_{r-1}=\mu_r\&\mu_2=\mu_{r-2}\&\cdots\&\mu_l=\mu_{l+1}, \]
where $r=2l+1$. Thus the level is given by
\begin{equation}
    k\equiv\mu_0+\mu_1+2\sum_{j=2}^{r-2}\mu_j+\mu_{r-1}+\mu_r=4\left(\mu_0+\sum_{j=2}^l\mu_j\right)\in4\mbb Z,\label{Doddk}
\end{equation}
reproducing the anomaly-free condition $k\in4\mbb Z$.

\subsection{$E_6$}
In this case, $A$ maps
\[ A[\mu_0;\mu_1,\mu_2,\mu_3,\mu_4,\mu_5,\mu_6]=[\mu_5;\mu_0,\mu_6,\mu_3,\mu_2,\mu_1,\mu_4]. \]
$A\hat\mu=\hat\mu$ imposes
\[ \mu_0=\mu_1=\mu_5\&\mu_2=\mu_4=\mu_6\&\mu_3\text{ is free}. \]
Thus the level is given by
\begin{equation}
    k\equiv\mu_0+\mu_1+2\mu_2+3\mu_3+2\mu_4+\mu_5+2\mu_6=3\left(\mu_0+2\mu_2+\mu_3\right)\in3\mbb Z,\label{E6k}
\end{equation}
reproducing the anomaly-free condition $k\in3\mbb Z$. $h^2$ or $h^{-1}$ clearly give the same condition.

\subsection{$E_7$}
In this case, we have
\[ A[\mu_0;\mu_1,\mu_2,\mu_3,\mu_4,\mu_5,\mu_6,\mu_7]=[\mu_6;\mu_5,\mu_4,\mu_3,\mu_2,\mu_1,\mu_0,\mu_7]. \]
$A\hat\mu=\hat\mu$ imposes
\[ \mu_0=\mu_6\&\mu_1=\mu_5\&\mu_2=\mu_4\&\mu_3,\mu_7\text{ are free}. \]
Thus the level is given by
\begin{equation}
    k\equiv\mu_0+2\mu_1+3\mu_2+4\mu_3+3\mu_4+2\mu_5+\mu_6+2\mu_7=2\left(\mu_0+2\mu_1+3\mu_2+2\mu_3+\mu_7\right)\in2\mbb Z,\label{E7k}
\end{equation}
reproducing the anomaly-free condition $k\in2\mbb Z$.

We summarize our results in Table \ref{table}.
\begin{table}[h]\label{table}
\begin{center}
\begin{tabular}{c|c|c|c|c}
type&center $\Gamma$&CS$_3$&$SZ_{(h,h)}\Big|=Z_{(h,h)}\Big|$&$|A\hat\mu\ra_c=|\hat\mu\ra_c$\\\hline
$A_r$&$\mbb Z_{r+1}$&$k\in(r+1)\mbb Z$&$k\in(r+1)\mbb Z$&$k\in(r+1)\mbb Z$\\
$B_r$&$\mbb Z_2$&$k\in\mbb Z$&$k\in\mbb Z$&$k\in\mbb Z$\\
$C_r$&$\mbb Z_2$&$rk\in2\mbb Z$&$rk\in2\mbb Z$&$rk\in2\mbb Z$\\
$D_{2l}$&$\mbb Z_2\times\mbb Z_2$&$k\in2\mbb Z$&$lk\in2\mbb Z$&$k\in2\mbb Z$\\
$D_{2l+1}$&$\mbb Z_4$&$k\in4\mbb Z$&$k\in4\mbb Z$&$k\in4\mbb Z$\\
$E_6$&$\mbb Z_3$&&$k\in3\mbb Z$&$k\in3\mbb Z$\\
$E_7$&$\mbb Z_2$&&$k\in2\mbb Z$&$k\in2\mbb Z$
\end{tabular}
\caption{Anomaly-free conditions}\label{summary}
\end{center}
\end{table}

%Note that all results are consistent with the cohomological classification (see e.g. \cite{CGLW})
%\[ H^3(\mbb Z_N,U(1))=\mbb Z_N,\quad H^3(\mbb Z_2\times\mbb Z_2,U(1))=\mbb Z_2\times\mbb Z_2\times\mbb Z_2. \]
We would like to make a few comments. 
%One may think the result $rk\in2\mbb Z$ of $C_r$ type does not fit the cohomological classification, but since in a given theory with a fixed rank and level, the anomaly is either ``on'' or ``off.'' So the result does fit the classification. Secondly, 
One would notice the ``mismatch'' in $D_{2l}$ type.\footnote{In this case, one can turn on discrete torsion \cite{torsion}. The possibility is discussed in \cite{KK}.} 
%The ``mismatch'' is originating from the mixed anomaly between $\mbb Z_2^A$ and $\mbb Z_2^{\wt A}$. (When and only when this mixed anomaly exists, there is another mixed anomaly between $\mbb Z_2^A\times\mbb Z_2^{\wt A}$ and large diffeomorphism \cite{KK}.) 
All twisted partition functions one can compute in the conventional formulation \cite{CFT} is of the form
\[ Z_{(h^l,h)} \]
where $h\in\mbb Z_N$ and $l=0,1,\dots,N-1$. That is why we have so far only computed ``diagonally twisted partition functions'' $Z_{(h,h)}$. However, using the generalized formalism \cite{KK}, one can also compute twisted partition functions including ``nondiagonally twisted partition functions'' $Z_{(h_t,h_x)}$. For example, in the case of $SO(4l)_k$ WZW models, such ``nondiagonally twisted partition functions" are given by
\begin{equation}
\begin{split}
    Z_{(h,\wt h)}&=\sum_{\hat\mu\in P^k_+}\bar\chi_{\wt A\hat\mu}\chi_{\hat\mu}e^{-2\pi i(A\hat\omega_0,\hat\mu+\frac k2\wt A\hat\omega_0)},\\
    Z_{(\wt h,h)}&=\sum_{\hat\mu\in P^k_+}\bar\chi_{A\hat\mu}\chi_{\hat\mu}e^{-2\pi i(\wt A\hat\omega_0,\hat\mu+\frac k2A\hat\omega_0)}.
\end{split}\label{SO(4l)_knondiagtwist}
\end{equation}
Performing the modular $S$-transformation on the first twisted partition function $Z_{(h,\wt h)}$, one obtains
\[ SZ_{(h,\wt h)}=\sum_{\hat\mu\in P^k_+}\bar\chi_{A\hat\mu}\chi_{\hat\mu}e^{-2\pi i(\wt A\hat\omega_0,A\hat\mu+\frac k2A\hat\omega_0)}. \]
Since we are considering $H=\mbb Z_2^A\times\mbb Z_2^{\wt A}$, which is larger than $\mbb Z_2$, our criterion requires to truncate the partition function to contributions coming just from the generators of $H$. Then the truncated twisted partition function is given by
\[ SZ_{(h,\wt h)}\Big|=\bar\chi_{k\hat\omega_1}\chi_{k\hat\omega_0}e^{-2\pi i(k/4+k/2)}+\bar\chi_{k\hat\omega_0}\chi_{k\hat\omega_1}e^{-2\pi ik/4}+\bar\chi_{k\hat\omega_r}\chi_{k\hat\omega_{r-1}}e^{-2\pi i(k/4+lk/2)}+\bar\chi_{k\hat\omega_{r-1}}\chi_{k\hat\omega_r}e^{-2\pi i(k/4+(l-1)k/2)}, \]
or in the matrix form
\begin{equation}
    SZ_{(h,\wt h)}\Big|\lr e^{-2\pi ik/4}\begin{pmatrix}0&1&0&0\\e^{-2\pi ik/2}&0&0&0\\0&0&0&e^{-2\pi ilk/2}\\0&0&e^{-2\pi i(l-1)k/2}&0\end{pmatrix}=:U'\label{SO(4l)_knondiagSZmatrix}
\end{equation}
in the basis $\{k\hat\omega_0,k\hat\omega_1,k\hat\omega_{r-1},k\hat\omega_r\}$, where as before row and column label $\chi$ and $\bar\chi$, respectively. Similarly, truncating the second twisted partition function of (\ref{SO(4l)_knondiagtwist}), one obtains
\[ Z_{(\wt h,h)}\Big|=\bar\chi_{k\hat\omega_1}\chi_{k\hat\omega_0}e^{-2\pi ik/4}+\bar\chi_{k\hat\omega_0}\chi_{k\hat\omega_1}e^{-2\pi i(k/4+k/2)}+\bar\chi_{k\hat\omega_r}\chi_{k\hat\omega_{r-1}}e^{-2\pi i(k/4+(l-1)k/2)}+\bar\chi_{k\hat\omega_{r-1}}\chi_{k\hat\omega_r}e^{-2\pi i(k/4+lk/2)}, \]
or in the matrix form
\begin{equation}
    Z_{(\wt h,h)}\Big|\lr e^{-2\pi ik/4}\begin{pmatrix}0&e^{-2\pi ik/2}&0&0\\1&0&0&0\\0&0&0&e^{-2\pi i(l-1)k/2}\\0&0&e^{-2\pi ilk/2}&0\end{pmatrix}=:U.\label{SO(4l)_knondiagZmatrix}
\end{equation}
Thus the $\wt D$ matrix is given by
\begin{equation}
    \wt D:=U^{-1}U'=\begin{pmatrix}e^{-2\pi ik/2}&&&\\&e^{2\pi ik/2}&&\\&&e^{-2\pi ik/2}&\\&&&e^{2\pi ik/2}\end{pmatrix}.\label{SO(4l)_ktildeDmatrix}
\end{equation}
The $\wt D$ matrix is equal to the identity matrix iff $k\in2\mbb Z$, indicating that there is no mixed anomaly between $\mbb Z_2^A$ and $\mbb Z_2^{\wt A}$ iff $k\in2\mbb Z$. Provided this result, the results of $D_{2l}$ type in the table \ref{summary} match with the reduced density matrix method as well as invariant boundary state criterion. 
%Although we believe our criterion
%\[ SZ_{(h,h')}\Big|=Z_{(h',h)}\Big| \]
%also works when $h$ and $h'$ are different, unfortunately we do not know how to compute $Z_{(h,h')}$.\footnote{Using the generalized formalism, one can also compute the twisted partition function. In case of $SO(4l)_k$ WZW models, they are given by
%Performing the modular $S$-transformation on the first twisted partition function, one obtains two equivalent expressions
%\begin{equation}
%    SZ_{(h,\wt h)}=\sum_{\hat\mu\in P^k_+}\chi_{A\hat\mu}\bar\chi_{\hat\mu}e^{-2\pi i(\wt A\hat\omega_0,A\hat\mu+\frac k2\wt A\hat\omega_0)}=\sum_{\hat\mu\in P^k_+}\chi_{A^{-1}\hat\mu}\bar\chi_{\hat\mu}e^{-2\pi i(\wt A\hat\omega_0,A^{-1}\hat\mu+\frac k2A\hat\omega_0)}.\label{SZhtildeh}
%\end{equation}
%One would notice that they are different from $Z_{(\wt h,h)}$ suggesting an existence of anomaly, but the mismatch is independent of the level $k$, and it seems that $SZ=Z$ does not work. (Since we are studying $\mbb Z_2^A\times\mbb Z_2^{\wt A}$ which is larger than $\mbb Z_2$, truncation may resolve the problem.)} That is why our results in the column is restricted to the case $h=h'$. However, as we saw in many examples,
%\[ SZ_{(h,h)}\Big|=Z_{(h,h)}\Big| \]
%correctly detect the braiding of $h$ with itself. So what is missing is only the braiding of $h$ with $\wt h$. As we show below, an existence of an invariant boundary state rules out anomalies including mixed ones.
Combining all these results, one can identify the anomaly which exists when $l\in2\mbb Z$ and $k\in2\mbb Z+1$ as a purely mixed anomaly between $\mbb Z_2^A$ and $\mbb Z_2^{\wt A}$. In this way, one can gain more detailed information about anomalies. Furthermore, this example supports more general version of our criterion (\ref{criteria0}). Finally, the result proves the equivalence
\begin{equation}
    \text{edgeable with }\Gamma\iff\Gamma\text{ is anomaly free}\label{edgecohomologyGamma}
\end{equation}
when one uses the full center $\Gamma$. As we explained in the introduction (\ref{edgecohomology'}), $(\Rightarrow)$ is automatic from the nilpotence of the boundary operator $\partial^2=0$. It turns out that the opposite direction $(\Leftarrow)$ does not hold in general when one considers subgroups of the centers as we see below, and it seems accidental that this direction holds when one considers the full center $\Gamma$.

\subsection{Invariant boundary states and anomaly decoupling}\label{decoupling}
After studying many examples in previous subsections, we would like to discuss the general relation between anomalies and invariant boundary states. Recently it has been conjectured that an $H$-invariant boundary state will indicate that $H$ is ``anomaly-decoupled''~\cite{Y19}. More precisely, consider an arbitrary subgroup $H$ of the entire global symmetry. If there exists an $H$-invariant boundary state, then $H$ is completely free of anomalies, including both anomalies involving only $H$ and mixed anomalies between $H$ and other symmetries. We prove in this section that this is indeed true in WZW models, which therefore supports the conjecture. 

Let us denote the generators of $H$ by $A_i$ $(i\in I)$.\footnote{When there is only one generator, we will omit the subscript for notational economy.} Thus if $H=\mbb Z_N$, $A^N=id_H$, and if $H=\mbb Z_M\times\mbb Z_N$, $I=\{1,2\}$ and $A_1^M=id_H,A_2^N=id_H$. We would like to show
\[ \exists\hat\mu\in P^k_+\ s.t.\ \forall i\in I,A_i\hat\mu=\hat\mu\iff\forall A'\in\mcal O(\hat{\text{g}}),\forall i\in I,\ e^{-2\pi ik(A_i\hat\omega_0,A'\hat\omega_0)}=1\ ,\]
where the equation in the right hand side means that the group $H$ constructed of $A_i$s is decoupled with other symmetries. To support the conjecture in~\cite{Y19}, i.e., the existence of an invariant boundary state implies anomaly-decoupling, we only need to show $(\Rightarrow)$. We postpone the discussion of the opposite direction $(\Leftarrow)$ to appendix \ref{invanom}.

$(\Rightarrow)$ We first consider the case $H=B(G)$. Recall that for any element $b\in B(G)$, there corresponds an element $A\in\mcal O(\hat{\text{g}})$ via
\begin{equation}{b\hat\lambda = \hat\lambda~b(\hat\lambda) = \hat\lambda~e^{-2\pi i (A\hat\omega_0,\hat\lambda)}}\ .\end{equation} Consider the commutation of $b_i$ with another element of outer automorphism $A'$, in the case of WZW models we have
\begin{equation}\label{commutation} b_iA'=A'b_ie^{-2\pi ik(A_i\hat\omega_0,A'\hat\omega_0)}\ , \end{equation}
which is eq.(17.31) in~\cite{CFT}. Given the invariant boundary state characterized by an affine weight $\hat\mu\in P^k_+$, consider the action of $b_iA'$ on $\hat\mu$
\begin{align*}
    b_iA'\hat\mu%&=b_i(A'\hat\mu)\\
    %&=e^{-2\pi i(A_i\hat\omega_0,A'\hat\mu)}A'\hat\mu\\
    &=A'b_ie^{-2\pi ik(A_i\hat\omega_0,A'\hat\omega_0)}\hat\mu\\
    &=e^{-2\pi ik(A_i\hat\omega_0,A'\hat\omega_0)}A'b_i\hat\mu\\
    &=e^{-2\pi ik(A_i\hat\omega_0,A'\hat\omega_0)}e^{-2\pi i(A_i\hat\omega_0,\hat\mu)}A'\hat\mu,
\end{align*}
where we used (\ref{commutation}).
% the general formula
%\[ b_iA'=A'b_ie^{-2\pi ik(A_i\hat\omega_0,A'\hat\omega_0)} \]
%for the center element $b_i\in B(G)$ corresponding to $A_i\in\mcal O(\hat{\text{g}})$.
\footnote{Since $H$ is a subgroup of the center, this is also trivially true for $b_i\in H$.}
%Since $A_j\hat\mu=\hat\mu$ for any $j\in J$ by assumption, we can raise $A_i$ to any power $q\in\mbb N$; $A_i^q\hat\mu=\hat\mu$. 
Since $A'$ is an element of $\mcal O(\hat{\text{g}})\simeq H$, there exists $p',q',\dots\in\mbb N$ such that $A'=A_1^{p'}A_2^{q'}\cdots$. Thus we have
\[ A'\hat\mu=A_1^{p'}A_2^{q'}\cdots\hat\mu=\hat\mu\ , \]
where we have used $A_i^q\hat\mu=\hat\mu$ which is a consequence of invariant boundary state condition.
Plugging this into $ b_iA'\hat\mu$ just computed,  we obtain 
\[ 0=\Big(e^{-2\pi ik(A_i\hat\omega_0,A'\hat\omega_0)}-1\Big)e^{-2\pi i(A_i\hat\omega_0,\hat\mu)}\hat\mu\ . \]
Multiplying both sides by $e^{2\pi i(A_i\hat\omega_0,\hat\mu)}$ we arrive
\[ 0=\Big(e^{-2\pi ik(A_i\hat\omega_0,A'\hat\omega_0)}-1\Big)\hat\mu. \]
%Since $\hat\mu=\mu_0\hat\omega_0+\mu_1\hat\omega_1+\cdots+\mu_r\hat\omega_r$, and using the linear independence of $\hat\omega_j$ $(j=0,1,\dots,r)$ we can rewrite this as
%\[ \Big(e^{-2\pi ik(A_i\hat\omega_0,A'\hat\omega_0)}-1\Big)\mu_0=\Big(e^{-2\pi ik(A_i\hat\omega_0,A'\hat\omega_0)}-1\Big)\mu_1=\cdots=\Big(e^{-2\pi ik(A_i\hat\omega_0,A'\hat\omega_0)}-1\Big)\mu_r=0. \]
%Since $\mu_j$s are in general nonzero, and the overall factor $\Big(e^{-2\pi ik(A_i\hat\omega_0,A'\hat\omega_0)}-1\Big)$ is $\hat\mu$ independent, 
When $\hat\mu$ is nonzero, which is true in our discussion, we must have
\[ e^{-2\pi ik(A_i\hat\omega_0,A'\hat\omega_0)}=1, \]
proving the desired result.

Next, let us consider subgroups $H$. Only $A_r$ and $D_r$ type have nontrivial subgroups, so we focus on them.

\underline{$A_r$ type}~~ In this case, $G=\mbb Z_{r+1}$. Suppose $r+1=lm$ where $l$ and $m$ are integers larger than one. Then there is a nontrivial subgroup $\mbb Z_l\subset\mbb Z_{lm}$.\footnote{One can of course repeat the following argument exchanging $l$ and $m$. Without loss of generality, we only consider the subgroup $\mbb Z_l$ below.} Denoting the generator of $\mbb Z_{lm}$ as $A$, i.e., $A^{lm}=id_G$, the subgroup is generated by $A^m$, $\left(A^m\right)^l=id_G$. Assume there exists $\hat\mu\in P^k_+$ such that $A^m\hat\mu=\hat\mu$. Then it requires
\begin{align*}
    \mu_0=\mu_m=\mu_{2m}=\cdots=\mu_{(l-1)m}&\&\mu_1=\mu_{m+1}=\mu_{2m+1}\cdots=\mu_{(l-1)m+1}\&\\
    &\cdots\&\mu_{m-1}=\mu_{2m-1}=\mu_{3m-1}=\cdots=\mu_{lm-1}.
\end{align*}
Thus the level is given by
\[ k\equiv\sum_{j=0}^{lm-1}\mu_j=l\left(\mu_0+\mu_1+\cdots+\mu_{m-1}\right)\in l\mbb Z. \]
Now, we would like to show that if $k$ is a multiple of $l$, then the phases
\[ e^{-2\pi ik(A^m\hat\omega_0,A'\hat\omega_0)}\]
 are trivial. The scalar products are computed as
\begin{align*}
    (A^m\hat\omega_0,A^j\hat\omega_0)&=F_{mj}=j\frac{l-1}l\quad(j=0,1,\dots,m),\\
    (A^m\hat\omega_0,A^{j'}\hat\omega_0)&=F_{mj'}=\frac{lm-j'}l\quad(j'=m+1,m+2,\dots,lm-1).
 \end{align*}
 Therefore, if $k\in l\mbb Z$, $k(A^m\hat\omega_0,A'\hat\omega_0)\in\mbb Z$. We conclude
 \[ \forall A'\in\mcal O(\hat{\text{g}}),\quad e^{-2\pi ik(A^m\hat\omega_0,A'\hat\omega_0)}=1. \]

\underline{$D_{2l}$ type}~~ In this case, $G=\mbb Z_2^A\times\mbb Z_2^{\wt A}$, and there are three nontrivial subgroups $H=\mbb Z_2^A,\mbb Z_2^{\wt A},\mbb Z_2^{\wt AA}$, where we denote different generators by superscripts. Let us study each subgroup in turn.

\begin{itemize}
\item $\mbb Z_2^A$: In this case, as we saw before, $A\hat\mu=\hat\mu$ requires
\[ \mu_0=\mu_1\&\mu_{r-1}=\mu_r. \]
So the level is given by
\[ k\equiv\mu_0+\mu_1+2\sum_{j=2}^{r-2}\mu_j+\mu_{r-1}+\mu_r=2\left(\mu_0+\sum_{j=2}^{r-2}\mu_j+\mu_{r-1}\right)\in2\mbb Z. \]
The scalar products appearing in the phase $e^{-2\pi ik(A\hat\omega_0,A'\hat\omega_0)}$ are computed as
\begin{align*}
    (A\hat\omega_0,\hat\omega_0)&=0,\\
    (A\hat\omega_0,A\hat\omega_0)&=(\hat\omega_1,\hat\omega_1)=1,\\
    (A\hat\omega_0,\wt A\hat\omega_0)&=(\hat\omega_1,\hat\omega_r)=\frac12,\\
    (A\hat\omega_0,\wt AA\hat\omega_0)&=(\hat\omega_1,\hat\omega_{r-1})=\frac12.
\end{align*}
Thus if $k$ is even, $k(A\hat\omega_0,A'\hat\omega_0)\in\mbb Z$, and we have
\[ \forall A'\in\mcal O(\hat{\text{g}}),\quad e^{-2\pi ik(A\hat\omega_0,A'\hat\omega_0)}=1. \]

\item $\mbb Z_2^{\wt A}$: In this case, as we saw before, $\wt A\hat\mu=\hat\mu$ requires
\[ \mu_0=\mu_r\&\mu_1=\mu_{r-1}\&\cdots\&\mu_{l-1}=\mu_{l+1}\&\mu_l\text{ is free}. \]
So the level is given by
\[ k\equiv\mu_0+\mu_1+2\sum_{j=2}^{r-2}\mu_j+\mu_{r-1}+\mu_r=2\left(\mu_0+\mu_1+2\sum_{j=2}^{l-1}\mu_j+\mu_l\right)\in2\mbb Z. \]
The scalar products appearing in the phase $e^{-2\pi ik(\wt A\hat\omega_0,A'\hat\omega_0)}$ are computed as
\begin{align*}
    (\wt A\hat\omega_0,\hat\omega_0)&=0,\\
    (\wt A\hat\omega_0,A\hat\omega_0)&=(\hat\omega_r,\hat\omega_1)=\frac12,\\
    (\wt A\hat\omega_0,\wt A\hat\omega_0)&=(\hat\omega_r,\hat\omega_r)=\frac l2,\\
    (\wt A\hat\omega_0,\wt AA\hat\omega_0)&=(\hat\omega_r,\hat\omega_{r-1})=\frac{l-1}2.
\end{align*}
Thus if $k$ is even, $k(\wt A\hat\omega_0,A'\hat\omega_0)\in\mbb Z$, and we have
\[ \forall A'\in\mcal O(\hat{\text{g}}),\quad e^{-2\pi ik(\wt A\hat\omega_0,A'\hat\omega_0)}=1. \]

\item $\mbb Z_2^{\wt AA}$: In this case, $\wt AA\hat\mu=\hat\mu$ requires
\[ \mu_0=\mu_{r-1}\&\mu_1=\mu_r\&\cdots\&\mu_{l-1}=\mu_{l+1}\&\mu_l\text{ is free}. \]
So the level is given by
\[ k\equiv\mu_0+\mu_1+2\sum_{j=2}^{r-2}\mu_j+\mu_{r-1}+\mu_r=2\left(\mu_0+\mu_1+2\sum_{j=2}^{l-1}\mu_j+\mu_l\right)\in2\mbb Z. \]
The scalar products appearing in the phase $e^{-2\pi ik(A\hat\omega_0,A'\hat\omega_0)}$ are computed as
\begin{align*}
    (\wt AA\hat\omega_0,\hat\omega_0)&=0,\\
    (\wt AA\hat\omega_0,A\hat\omega_0)&=(\hat\omega_{r-1},\hat\omega_1)=\frac12,\\
    (\wt AA\hat\omega_0,\wt A\hat\omega_0)&=(\hat\omega_{r-1},\hat\omega_r)=\frac{l-1}2,\\
    (\wt AA\hat\omega_0,\wt AA\hat\omega_0)&=(\hat\omega_{r-1},\hat\omega_{r-1})=\frac l2.
\end{align*}
Thus if $k$ is even, $k(\wt AA\hat\omega_0,A'\hat\omega_0)\in\mbb Z$, and we have
\[ \forall A'\in\mcal O(\hat{\text{g}}),\quad e^{-2\pi ik(\wt AA\hat\omega_0,A'\hat\omega_0)}=1. \]
\end{itemize}
In short, we saw the phases are trivial in all cases if there exists invariant boundary states, as stated.

\underline{$D_{2l+1}$ type}~~ In this case, since $G=\mbb Z_4$, the only nontrivial subgroup is $H=\mbb Z_2$. Denoting the generator of $\mbb Z_4$ by $A$, i.e., $A^4=id_G$, the generator of $H$ is given by $A^2$. Suppose there exists $\hat\mu\in P^k_+$ such that $A^2\hat\mu=\hat\mu$. Using the action of $A$ as we gave before, the assumption requires
\[ A^2[\mu_0;\mu_1,\mu_2,\cdots,\mu_{r-2},\mu_{r-1},\mu_r]=[\mu_1;\mu_0,\mu_2,\cdots,\mu_{r-2},\mu_r,\mu_{r-1}]\stackrel!=\hat\mu, \]
or
\[ \mu_0=\mu_1\&\mu_{r-1}=\mu_r. \]
Thus the level is given by
\begin{equation}
    k\equiv\mu_0+\mu_1+2\sum_{j=2}^{r-2}\mu_j+\mu_{r-1}+\mu_r=2\left(\mu_0+\sum_{j=2}^{r-2}\mu_j+\mu_{r-1}\right)\in2\mbb Z.\label{DevenZ2}
\end{equation}
We would like to show if $k\in2\mbb Z$, then the phases 
\[ e^{-2\pi i(A^2\hat\omega_0,A'\hat\omega_0)} \]
are trivial.
The scalar products are computed as
\begin{align*}
    (A^2\hat\omega_0,\hat\omega_0)&=0,\\
    (A^2\hat\omega_0,A\hat\omega_0)&=(\hat\omega_1,\hat\omega_r)=\frac12,\\
    (A^2\hat\omega_0,A^2\hat\omega_0)&=(\hat\omega_1,\hat\omega_1)=1,\\
    (A^2\hat\omega_0,A^3\hat\omega_0)&=(\hat\omega_1,\hat\omega_{r-1})=\frac12.
\end{align*}
Therefore, if $k$ is even, $k(A^2\hat\omega_0,A'\hat\omega_0)\in\mbb Z$, and we conclude
\[ \forall A'\in\mcal O(\hat{\text{g}}),\quad e^{-2\pi ik(A^2\hat\omega_0,A'\hat\omega_0)}=1, \]
as stated.
In short, our results in this section can be summarized as
\begin{equation}
    H(\subset G)\text{-edgeable}\iff H\text{ is anomaly-decoupled}\subset H\text{ is anomaly free}.\label{edgecohomology}
\end{equation}
We leave the proof of the opposite direction $(\Leftarrow)$ in appendix \ref{invanom}. Thus we demonstrated that in WZW models invariant boundary state condition and anomaly decoupled are equivalent.

\section{Discussions}\label{discussion}
In this paper we proposed a modular transformation approach to detect an anomaly for a discrete internal global symmetry $G$ in two-dimensional diagonal RCFTs based on twisted torus partition functions.
This was motivated by searching for the two-dimensional analogy of a pair of linked symmetry lines (Wilson loops) in three-dimensional Chern-Simons theory in the light of bulk/boundary correspondence. We have explicitly shown that a criterion
\begin{equation}\label{criteria3}
SZ_{(h,h')}\Big| = Z_{(h',h)}\Big|
\end{equation} exactly reproduces the anomaly-free conditions for a large class of CFT models. The underlying intuition for our criterion (\ref{criteria3}) is that $S$-transformation can flip the ordering of insertions of topological defect lines. This criterion can detect both anomaly of symmetry $G$ and the mixed anomaly between symmetries $G_1$ and $G_2$ where $h$ and $h'$ are generators, $h\in G_1$ and $h'\in G_2$. Using twisted torus partition function we also generalize the orbifolding condition in~\cite{NY,CFT} to the cases when $G$ is a product group. One can view our criterion (\ref{criteria3}) and orbifolding condition as consistency conditions coming from modular $S$-transformation and modular $T$-transformation respectively, which indicates that modular transformations play important roles in detecting anomalies.
By explicitly analyzing WZW models and minimal models, we demonstrate a chain of relations:
\[ H(\subset G)\text{-edgeable}\iff H\text{-anomaly decoupled}\subset H\text{-anomaly free}\subset H\text{-orbifoldable}. \]
We believe that this chain of relations is true for all symmetries captured by Verlinde lines in diagonal rational conformal field theories. The validity of them for more general symmetries and more general conformal field theories is not obvious and we leave it as an interesting future problem. It would also be interesting to consider RG flows between WZW models as in \cite{TS} matching the anomalies.

\section*{Acknowledgement}
We are grateful for helpful discussions with Yang Qi, Shu Heng Shao, Yuji Tachikawa, Yidun Wan, Chenjie Wang and Shimon Yankielowicz.

\appendix
\setcounter{section}{0}
\renewcommand{\thesection}{\Alph{section}}
\setcounter{equation}{0}
\renewcommand{\theequation}{\Alph{section}.\arabic{equation}}

\section{Generalized orbifolding}\label{Go}
We review a generalized orbifolding procedure, focusing on WZW models following~\cite{CFT}. The center of the gauge group of WZWs is $B(\hat G)$, which is isomorphic to the outer automorphism symmetry $\mcal O (\hat{\text{g}})$. The first step is to project the Hilbert space of the diagonal theory
\begin{equation}
Z(\tau) = \sum_{\hat\lambda\in P^k_+} \chi_{\hat\lambda}(\tau) \bar\chi_{\hat\lambda}(\bar\tau)
\end{equation}
 onto the $B(\hat G)$ invariant states. Let $b$ be a generator of a cyclic group $B(\hat G)$ of order $N$, $b^N=1$. The projection operator is
\begin{equation}
\mathcal{B} = {1\over N}\sum_{q=0}^{N-1} b^q\ .
\end{equation} The action of $b$ on characters is defined by
\begin{equation}
b\,\chi_{\hat\lambda} = \chi_{\hat\lambda} b(\hat\lambda)\ .
\end{equation}
The operator to generate the twisted sectors is given by
\begin{equation}
\mathcal{A} = \sum_{p=0}^{N-1} A^p
\end{equation}
 with 
 \begin{equation}
 A\,\chi_{\hat\lambda} = \chi_{A\hat\lambda}\ .
 \end{equation}
It acts on the $B(\hat G)$ projected invariant states. Here $A$ is an element of $\mcal O (\hat{\text{g}})$. If $A$ and $b$ commute, the candidate mass matrix $\mathcal{M}=\mathcal{A}\mathcal{B}$ will be modular invariant. However this is not the case because $Ab\neq bA$. To compensate the noncommutativity we define a improved product $\star$ by
\begin{equation}
A'\star b := A'b e^{-ik\pi(A\hat\omega_0,A'\hat\omega_0)}\ ,\quad b\star A' := bA' e^{ik\pi(A\hat\omega_0,A'\hat\omega_0)}\ ,
\end{equation} which satisfy the commutation relation
\begin{equation}
A'\star b = b\star A'\ .
\end{equation}
One can check that the improved mass matrix 
\begin{equation}
\widetilde{\mathcal{M}} = {1\over N}\sum_{p,q=0}^{N-1} A^p\star b^q
\end{equation} is modular invariant. However this does not always make sense unless the following consistency condition is satisfied: $\widetilde{\mathcal{M}}$ must be invariant under $b^q\to b^{q+N}$. This is the same condition as one can obtain from
\begin{equation}
Z_{(h^N,h)}= Z_{(1,h)}\ ,
\end{equation} with $h$ being the symmetry line corresponding to $b$. This leads to the condition~\cite{NY,CFT}
\begin{equation}
{Nk\over 2}|A\hat\omega_0|^2 \in \mathbb{Z}\ .
\end{equation}

\section{Center invariant boundary state and the mixed anomaly}\label{invanom}
In this appendix, we would like to complete the proof of the equivalence of an existence of invariant boundary states under the subgroups of centers and the trivial phases in the commutation relation of $b_i\in B(G)$ and $A'\in\mcal O(\hat{\text{g}})$. More precisely, we would like to show
\[ \exists\hat\mu\in P^k_+\ s.t.\ \forall i\in I,A_i\hat\mu=\hat\mu\iff\forall A'\in\mcal O(\hat{\text{g}}),\forall i\in I,\ e^{-2\pi ik(A_i\hat\omega_0,A'\hat\omega_0)}=1. \]
Since we have already shown $(\Rightarrow)$ in subsection \ref{decoupling}, we just show $(\Leftarrow)$ here.

$(\Leftarrow)$ This direction cannot be proven algebraically, and one has to study each case in detail. The key of the proof is that the conditions imposed on the levels by the trivial phases fix WZW models to those with invariant boundary states. More precisely, once the level is fixed to those required in section \ref{IB}, exhaustive nature of dominant weights $P^k_+$ guarantees an existence of invariant boundary states. As in section \ref{IB}, we first consider full groups, then later consider subgroups.

\subsection{$A_r$ type, i.e., $\mfrak{su}(r+1)$}
In this case, the outer automorphism group (and the center group) is isomorphic to a cyclic group $\mcal O(\hat{\text{g}})\simeq\mbb Z_{r+1}$. Let us consider the fundamental element $A\in\mcal O(\hat{\text{g}})$ which maps
\[ A[\mu_0;\mu_1,\cdots,\mu_r]=[\mu_r;\mu_0,\cdots,\mu_{r-1}]. \]
Then, since $A'\hat\omega_0$ runs all $\hat\omega_j$ with $j=0,1,\cdots,r$, the trivial phase condition imposes
\[ \forall j\in\{0,1,\cdots,r\},\quad e^{-2\pi ik(\hat\omega_1,\hat\omega_j)}=e^{-2\pi ik\frac{r-j+1}{r+1}}\stackrel!=1. \]
Thus we obtain
\begin{equation}
    k\in(r+1)\mbb Z.\label{Aphase}
\end{equation}
To show there exists an invariant boundary state, let us consider the smallest positive level, namely $k=r+1$. Generalization to the other levels are given in short.

Because of the exhaustive nature of dominant weights $P^k_+$, our theory $\mfrak{su}(r+1)_{r+1}$ necessarily contains an affine weight
\[ \hat\mu_\text{I.B.}^{(r+1)}=[1;1,\cdots,1]. \]
One can easily convince oneself that the boundary state corresponding to this affine weight is invariant under $A\in\mcal O(\hat{\text{g}})$. Thus we could find an invariant state.

The other anomaly-free cases $k=n(r+1)$ with $n\in\mbb Z$ can be explored immediately by multiplying $n$ to the invariant state we found;
\[ \hat\mu_\text{I.B.}^{\big(n(r+1)\big)}=n\hat\mu_\text{I.B.}^{(r+1)}=[n;n,\cdots,n]. \]
Again, it is easy to see the boundary state of $\mfrak{su}(r+1)_{n(r+1)}$ WZW corresponding to this affine weight is invariant under $A\in\mcal O(\hat{\text{g}})$.

\subsection{$B_r$ type, i.e., $\mfrak{so}(2r+1)$}
The outer automorphism group (and the center group) is isomorphic to a cyclic group $\mcal O(\hat{\text{g}})\simeq\mbb Z_2$. So the only nontrivial element $A\in\mcal O(\hat{\text{g}})$ sends
\[ A[\mu_0;\mu_1,\mu_2,\cdots,\mu_r]=[\mu_1;\mu_0,\mu_2,\cdots,\mu_r]. \]
The trivial phase condition gives
\[ e^{-2\pi ik(\hat\omega_1,\hat\omega_0)}\stackrel!=1\stackrel!=e^{-2\pi ik(\hat\omega_1,\hat\omega_1)}. \]
The first equality is automatically satisfied. Since $(\hat\omega_1,\hat\omega_1)=1$, the second imposes
\begin{equation}
    k\in\mbb Z.\label{Bphase}
\end{equation}
This condition is trivially satisfied. Let us pick an arbitrary integer $k\in\mbb Z$, and try to construct an invariant boundary state in our $\mfrak{so}(2r+1)_k$ WZW model. As we saw in section \ref{IB}, an existence of an invariant boundary state requires $\mu_0=\mu_1$. Thus, for example, a boundary state corresponding to an affine weight
\[ \hat\mu_\text{I.B.}^{(k)}=[0;0,\cdots,0,k] \]
is invariant under $A$. Thus we could construct an invariant boundary state.

\subsection{$C_r$ type, i.e., $\mfrak{sp}(2r)$}
The outer automorphism group (and the center group) is isomorphic to a cyclic group $\mcal O(\hat{\text{g}})\simeq\mbb Z_2$. So nontrivial conditions are coming just from the nontrivial element $A\in\mcal O(\hat{\text{g}})$. It sends
\[ A[\mu_0;\mu_1,\cdots,\mu_r]=[\mu_r;\mu_{r-1},\cdots,\mu_0]. \]
The trivial phase condition imposes
\[ e^{-2\pi ik(\hat\omega_r,\hat\omega_r)}=(-)^{rk}\stackrel!=1. \]
Thus we get
\begin{equation}
    rk\in2\mbb Z.\label{Cphase}
\end{equation}
We have to consider odd and even $r$ cases separately.

$\underline{r\in2\mbb Z+1}$~~ In this case, the level $k$ has to be even. Then the dominant weights $P^k_+$ must contain, say,
\[ \hat\mu_\text{I.B.}^{(k)}=[k/2;0,\cdots,0,k/2]. \]
One can easily check $A\hat\mu_\text{I.B.}^{(k)}=\hat\mu_\text{I.B.}^{(k)}$, showing an existence of an invariant bounary state.

$\underline{r\in2\mbb Z}$~~ In this case, the level can be an odd integer. If it is even, the affine weight we have just considered provides an invariant boundary state. If it is odd, the dominant weight $P^k_+$ must contain, say,
\[ \hat\mu_\text{I.B.}^{(k)}=[(k-1)/2,0,\cdots,0,1,0,\cdots,0,(k-1)/2], \]
where the nonzero component in the middle is $\mu_{r/2}$. Since this is invariant under $A$, the corresponding boundary state is invariant.

\subsection{$D_r$ type, i.e., $\mfrak{so}(2r)$}
We study even and odd $r$ separately.

$\underline{r\in2\mbb Z}$~~ In this case, the outer automorphism group (and the center group) is isomorphic to $\mcal O(\hat{\text{g}})\simeq\mbb Z_2^A\times\mbb Z_2^{\wt A}$, where the superscripts denote generators. They map
\begin{align*}
    A[\mu_0;\mu_1,\mu_2,\cdots,\mu_{r-2},\mu_{r-1},\mu_r]&=[\mu_1;\mu_0,\mu_2,\cdots,\mu_{r-2},\mu_r,\mu_{r-1}],\\
    \wt A[\mu_0;\mu_1,\mu_2,\cdots,\mu_{r-2},\mu_{r-1},\mu_r]&=[\mu_r;\mu_{r-1},\mu_{r-2},\cdots,\mu_2,\mu_1,\mu_0].
\end{align*}
The trivial phase condition requires
\[ e^{-2\pi ik(\hat\omega_1,\hat\omega_0)}=e^{-2\pi ik(\hat\omega_1,\hat\omega_1)}=e^{-2\pi ik(\hat\omega_1,\hat\omega_r)}=e^{-2\pi ik(\hat\omega_1,\hat\omega_{r-1})}\stackrel!=1, \]
and
\[ e^{-2\pi ik(\hat\omega_r,\hat\omega_0)}=e^{-2\pi ik(\hat\omega_r,\hat\omega_1)}=e^{-2\pi ik(\hat\omega_r,\hat\omega_r)}=e^{-2\pi ik(\hat\omega_r,\hat\omega_{r-1})}\stackrel!=1. \]
Using the quadratic form matrix, these reduce to
\[ (-)^k=(-)^{lk}=(-)^{(l-1)k}\stackrel!=1, \]
giving
\begin{equation}
    k\in2\mbb Z.\label{Devenphase}
\end{equation}
To show an existence of an invariant boundary state, we follow our familiar logic; pick $k=2$. Then we can explicitly find an invariant boundary state corresponding to, say,
\[ \hat\mu_\text{I.B.}^{(2)}=[0;0,1,0,\dots,0,1,0,0], \]
i.e., $\mu_2=1=\mu_{r-2}$ and all other affine Dynkin labels are zero. When $k=2n$ with $n\in\mbb Z$,
\[ \hat\mu_\text{I.B.}^{(2n)}=n\hat\mu_\text{I.B.}^{(2)}=[0;0,n,0,\dots,0,n,0,0] \]
gives, for example, an invariant boundary state.

$\underline{r\in2\mbb Z+1}$~~ The outer automorphism group (and the center group) is isomorphic to a cyclic group $\mcal O(\hat{\text{g}})\simeq\mbb Z_4$. Let us consider the fundamental element $A\in\mcal O(\hat{\text{g}})$ which maps
\[ A[\mu_0;\mu_1,\mu_2,\cdots,\mu_{r-2},\mu_{r-1},\mu_r]=[\mu_{r-1};\mu_r,\mu_{r-2},\cdots,\mu_2,\mu_1,\mu_0]. \]
The trivial phase condition requires
\[ e^{-2\pi ik(\hat\omega_r,\hat\omega_0)}=e^{-2\pi ik(\hat\omega_r,\hat\omega_r)}=e^{-2\pi ik(\hat\omega_r,\hat\omega_1)}=e^{-2\pi ik(\hat\omega_r,\hat\omega_{r-1})}\stackrel!=1. \]
Using the quadratic form matrix, these conditions reduce to
\[ (-)^{(2l+1)k/2}=(-)^k=(-)^{(2l-1)k/2}\stackrel!=1, \]
giving
\begin{equation}
    k\in4\mbb Z.\label{Doddphase}
\end{equation}
To show there exists an invariant boundary state, let us follow the strategy of the case $A_r$, namely pick the smallest positive level $k=4$. Then the dominant weights $P^k_+$ must contain, say,
\[ \hat\mu_\text{I.B.}^{(4)}=[1;1,0,\cdots,0,1,1], \]
which is invariant under $A$. Thus the corresponding boundary state is invariant. The other anomaly-free cases can be dealt with ease; in an $\mfrak{so}(4l+2)_{4n}$ WZW model with $n\in\mbb Z$, take for example
\[ \hat\mu_\text{I.B.}^{(4n)}=n\hat\mu_\text{I.B.}^{(4)}=[n;n,0,\cdots,0,n,n]. \]
This affine weight provides an invariant boundary state.

\subsection{$E_6$}
The outer automorphism group (and the center group) is isomorphic to a cyclic group $\mcal O(\hat{\text{g}})\simeq\mbb Z_3$. Let us consider the fundamental element $A\in\mcal O(\hat{\text{g}})$ which sends
\[ A[\mu_0;\mu_1,\mu_2,\mu_3,\mu_4,\mu_5,\mu_6]=[\mu_5;\mu_0,\mu_6,\mu_3,\mu_2,\mu_1,\mu_4]. \]
The trivial phase condition imposes
\[ e^{-2\pi ik(\hat\omega_1,\hat\omega_0)}=e^{-2\pi ik(\hat\omega_1,\hat\omega_1)}=e^{-2\pi ik(\hat\omega_1,\hat\omega_5)}\stackrel!=1. \]
Using the quadratic form matrix, one obtains
\[ e^{-2\pi ik/3}\stackrel!=1, \]
giving
\begin{equation}
    k\in3\mbb Z.\label{E6phase}
\end{equation}
As before, pick $k=3$. Then the dominant weights $P^k_+$ must contain, say,
\[ \hat\mu_\text{I.B.}^{(3)}=[1;1,0,0,0,1,0]. \]
One can easily see this is invariant under $A$, providing an invariant boundary state. For other cases with level $k=3n$ with $n\in\mbb Z$,
\[ \hat\mu_\text{I.B.}^{(3n)}=n\hat\mu_\text{I.B.}^{(3)}=[n;n,0,0,0,n,0] \]
gives an invariant boundary state.

\subsection{$E_7$}
The outer automorphism group (and the center group) is isomorphic to a cyclic group $\mcal O(\hat{\text{g}})\simeq\mbb Z_2$. Thus nontrivial conditions are coming just from the nontrivial element $A\in\mcal O(\hat{\text{g}})$ which maps
\[ A[\mu_0;\mu_1,\mu_2,\mu_3,\mu_4,\mu_5,\mu_6,\mu_7]=[\mu_6;\mu_5,\mu_4,\mu_3,\mu_2,\mu_1,\mu_0,\mu_7]. \]
The only nontrivial condition forced by the trivial phase condition is
\[ e^{-2\pi ik(\hat\omega_6,\hat\omega_6)}\stackrel!=1. \]
With the help of the quadratic form matrix, one obtains
\[ (-)^{3k}\stackrel!=1, \]
giving
\begin{equation}
    k\in2\mbb Z.\label{E7phase}
\end{equation}
As usual, pick $k=2$, then the dominant weights $P^k_+$ must contain, say,
\[ \hat\mu_\text{I.B.}^{(2)}=[1;0,0,0,0,0,1,0], \]
which is invariant under $A$. Thus the corresponding boundary state is invariant. For other cases with level $k=2n$ with $n\in\mbb Z$,
\[ \hat\mu_\text{I.B.}^{(2n)}=n\hat\mu_\text{I.B.}^{(2)}=[n;0,0,0,0,0,n,0] \]
provides an invariant boundary state.

\subsection{Subgroups}
Finally, let us consider subgroups. Since we already gave essential computations in subsection \ref{decoupling}, we would be rather brief in this subsection.

\underline{$A_r$ type}~~ As in subsection \ref{decoupling}, let us suppose $r+1=lm$, and consider a subgroup $\mbb Z_l\subset\mbb Z_{lm}$. Since the scalar products appearing in the phase $e^{-2\pi ik(A^m\hat\omega_0,A'\hat\omega_0)}$ are proportional to $1/l$, the trivial phase condition forces $k\in l\mbb Z$.\footnote{One may ask whether $k\in l'\mbb Z$ with $|l'|<|l|$ could give trivial phases. However, there does not exist such $l'$. The reason is simply because $F_{m,lm-1}=1/l$ and $l'/l$ cannot be an integer.} Since this condition is the same as required by an existence of invariant boundary state, it is guaranteed that there exists an invariant boundary state. As a demonstration, pick $k=l$. Then
\[ \hat\mu_\text{I.B.}^{(l)}=[1;0,0,\dots,0,1,0,0,\dots,0,\dots] \]
gives an invariant boundary state where $\mu_0=\mu_m=\mu_{2m}=\cdots=\mu_{(l-1)m}=1$ and all other affine Dynkin labels are zero. For $k=ln$ with $n\in\mbb Z$, $n\hat\mu_\text{I.B.}^{(l)}$ gives an invariant boundary state.

\underline{$D_{2l}$ type}~~ In subsection \ref{decoupling}, we saw the scalar products appearing in the phase $e^{-2\pi ik(A_i\hat\omega_0,A'\hat\omega_0)}$ have the form $n/2$ with some integer $n$.
Thus the trivial phase condition requires $k\in2\mbb Z$. Since this is the same condition as required by an existence of invariant boundary state for any subgroups, the former restricts the theory to WZW models with invariant boundary states. One can easily find explicit affine weights invariant under each subgroup following our usual argument.

\underline{$D_{2l+1}$ type}~~ The computation in subsection \ref{decoupling} showed the scalar products are half-integer valued. Thus the trivial phase condition requires $k\in2\mbb Z$. Because this condition is the same as that imposed by an existence of invariant boundary state, one cannot avoid having invariant boundary state once one set $k\in2\mbb Z$ due to the exhaustive nature of dominant weights $P^k_+$.

\section{$Ch$ invariant boundary states and orbifoldability}\label{Ch}
In this appendix, we show an equivalence of an existence of $Ch$ invariant boundary state and the orbifoldability. More precisely, we would like to show
\[ \exists\hat\mu\in P^k_+\ s.t.\forall i\in I,\ CA_i\hat\mu=\hat\mu\iff\text{consistency condition originating from modular }T\text{-transformations}. \]
As noticed in \cite{NY}, when $H=\mbb Z_N$, the orbifoldability condition is given by
\[ \frac{kN}2|A\hat\omega_0|^2\in\mbb Z, \]
where $A$ is the generator of $\mbb Z_N$, i.e., $A^N=id_H$. In this case, the equivalence was already shown in \cite{NY}, so we only have to show the equivalence for $D_{2l}$. We note that this equivalence only holds for $H=\mbb Z_2\times\mbb Z_2$ for the case.

$(\Leftarrow)$ In this case, the consistency condition is given by \cite{KK}
\begin{equation}
    k\in2\mbb Z.\label{Devenconsistency}
\end{equation}
To show an existence of an invariant boundary state, we follow our familiar argument; pick $k=2$. Since the charge conjugation $C$ is trivial in this case, an affine weight
\[ \hat\mu_\text{I.B.}^{(2)}=[0;0,1,0,\cdots,0,1,0,0], \]
that is $\mu_2=1=\mu_{r-2}$ and all other affine Dynkin labels are zero, provides an invariant boundary state. In fact the affine weight is invariant under both $A$ and $\wt A$, thus under the whole $H=\mbb Z_2^A\times\mbb Z_2^{\wt A}$. For $k=2n$ with $n\in\mbb Z$, $n\hat\mu_\text{I.B.}^{(2)}$ gives an invariant boundary state. Therefore the orbfoldability condition guarantees an existence of an invariant boundary state, as stated.

$(\Rightarrow)$ As we computed in (\ref{Devenk}), an existence of invariant boundary state under both $A$ and $\wt A$ requires $k$ be even. This is the same as (\ref{Devenconsistency}).


\begin{thebibliography}{30}
\bibitem{H80}
  G.~'t Hooft,
  ``Naturalness, chiral symmetry, and spontaneous chiral symmetry breaking,''
  NATO Sci.\ Ser.\ B {\bf 59}, 135 (1980).
  doi:10.1007/978-1-4684-7571-5$\_$9
  %%CITATION = doi:10.1007/978-1-4684-7571-5_9;%%
  %451 citations counted in INSPIRE as of 13 Jun 2019
\bibitem{GKKS}
  D.~Gaiotto, A.~Kapustin, Z.~Komargodski and N.~Seiberg,
  ``Theta, Time Reversal, and Temperature,''
  JHEP {\bf 1705}, 091 (2017)
  doi:10.1007/JHEP05(2017)091
  [arXiv:1703.00501 [hep-th]].
  %%CITATION = doi:10.1007/JHEP05(2017)091;%%
  %95 citations counted in INSPIRE as of 19 Jun 2019
\bibitem{JSY}
  K.~Jensen, E.~Shaverin and A.~Yarom,
  ``’t Hooft anomalies and boundaries,''
  JHEP {\bf 1801}, 085 (2018)
  doi:10.1007/JHEP01(2018)085
  [arXiv:1710.07299 [hep-th]].
  %%CITATION = doi:10.1007/JHEP01(2018)085;%%
  %9 citations counted in INSPIRE as of 25 May 2019
\bibitem{Gaiotto:2017tne}
  D.~Gaiotto, Z.~Komargodski and N.~Seiberg,
  ``Time-reversal breaking in QCD$_{4}$, walls, and dualities in 2 + 1 dimensions,''
  JHEP {\bf 1801}, 110 (2018)
  doi:10.1007/JHEP01(2018)110
  [arXiv:1708.06806 [hep-th]].
  %%CITATION = doi:10.1007/JHEP01(2018)110;%%
  %71 citations counted in INSPIRE as of 17 Jun 2019
\bibitem{CGLW}
  X.~Chen, Z.~C.~Gu, Z.~X.~Liu and X.~G.~Wen,
  ``Symmetry protected topological orders and the group cohomology of their symmetry group,''
  Phys.\ Rev.\ B {\bf 87}, no. 15, 155114 (2013)
  doi:10.1103/PhysRevB.87.155114
  [arXiv:1106.4772 [cond-mat.str-el]].
  %%CITATION = doi:10.1103/PhysRevB.87.155114;%%
  %381 citations counted in INSPIRE as of 16 Feb 2019
\bibitem{Hung:2012nf}
  L.~Y.~Hung and X.~G.~Wen,
  ``Quantized topological terms in weak-coupling gauge theories with a global symmetry and their connection to symmetry-enriched topological phases,''
  Phys.\ Rev.\ B {\bf 87}, no. 16, 165107 (2013)
  doi:10.1103/PhysRevB.87.165107
  [arXiv:1212.1827 [cond-mat.str-el]].
  %%CITATION = doi:10.1103/PhysRevB.87.165107;%%
  %61 citations counted in INSPIRE as of 17 Jun 2019
\bibitem{WZ71}
  J.~Wess and B.~Zumino,
  ``Consequences of anomalous Ward identities,''
  Phys.\ Lett.\  {\bf 37B}, 95 (1971).
  doi:10.1016/0370-2693(71)90582-X
  %%CITATION = doi:10.1016/0370-2693(71)90582-X;%%
  %2607 citations counted in INSPIRE as of 13 Jun 2019
\bibitem{FS}
  L.~D.~Faddeev and S.~L.~Shatashvili,
  ``Algebraic and Hamiltonian Methods in the Theory of Nonabelian Anomalies,''
  Theor.\ Math.\ Phys.\  {\bf 60}, 770 (1985)
  [Teor.\ Mat.\ Fiz.\  {\bf 60}, 206 (1984)].
  doi:10.1007/BF01018976
  %%CITATION = doi:10.1007/BF01018976;%%
  %164 citations counted in INSPIRE as of 25 May 2019
\bibitem{CH}
  C.~G.~Callan, Jr. and J.~A.~Harvey,
  ``Anomalies and Fermion Zero Modes on Strings and Domain Walls,''
  Nucl.\ Phys.\ B {\bf 250}, 427 (1985).
  doi:10.1016/0550-3213(85)90489-4
  %%CITATION = doi:10.1016/0550-3213(85)90489-4;%%
  %513 citations counted in INSPIRE as of 25 May 2019
\bibitem{KT}
  A.~Kapustin and R.~Thorngren,
  ``Anomalies of discrete symmetries in various dimensions and group cohomology,''
  arXiv:1404.3230 [hep-th].
  %%CITATION = ARXIV:1404.3230;%%
  %86 citations counted in INSPIRE as of 29 May 2019
\bibitem{GKSW}
  D.~Gaiotto, A.~Kapustin, N.~Seiberg and B.~Willett,
  ``Generalized Global Symmetries,''
  JHEP {\bf 1502}, 172 (2015)
  doi:10.1007/JHEP02(2015)172
  [arXiv:1412.5148 [hep-th]].
  %%CITATION = doi:10.1007/JHEP02(2015)172;%%
  %196 citations counted in INSPIRE as of 24 May 2019
\bibitem{CLSWY}
  C.~M.~Chang, Y.~H.~Lin, S.~H.~Shao, Y.~Wang and X.~Yin,
  ``Topological Defect Lines and Renormalization Group Flows in Two Dimensions,''
  JHEP {\bf 1901}, 026 (2019)
  doi:10.1007/JHEP01(2019)026
  [arXiv:1802.04445 [hep-th]].
  %%CITATION = doi:10.1007/JHEP01(2019)026;%%
  %6 citations counted in INSPIRE as of 29 May 2019
\bibitem{BT}
  L.~Bhardwaj and Y.~Tachikawa,
  ``On finite symmetries and their gauging in two dimensions,''
  JHEP {\bf 1803}, 189 (2018)
  doi:10.1007/JHEP03(2018)189
  [arXiv:1704.02330 [hep-th]].
  %%CITATION = doi:10.1007/JHEP03(2018)189;%%
  %8 citations counted in INSPIRE as of 29 May 2019
\bibitem{2group}
  C.~Cordova, T.~T.~Dumitrescu and K.~Intriligator,
  ``Exploring 2-Group Global Symmetries,''
  JHEP {\bf 1902}, 184 (2019)
  doi:10.1007/JHEP02(2019)184
  [arXiv:1802.04790 [hep-th]];
  %%CITATION = doi:10.1007/JHEP02(2019)184;%%
  %33 citations counted in INSPIRE as of 29 May 2019
  F.~Benini, C.~Cordova and P.~S.~Hsin,
  ``On 2-Group Global Symmetries and their Anomalies,''
  JHEP {\bf 1903}, 118 (2019)
  doi:10.1007/JHEP03(2019)118
  [arXiv:1803.09336 [hep-th]].
  %%CITATION = doi:10.1007/JHEP03(2019)118;%%
  %24 citations counted in INSPIRE as of 29 May 2019
  
  %\cite{Harlow:2018tng}
\bibitem{Harlow:2018tng} 
  D.~Harlow and H.~Ooguri,
  ``Symmetries in quantum field theory and quantum gravity,''
  arXiv:1810.05338 [hep-th].
  %%CITATION = ARXIV:1810.05338;%%
  %39 citations counted in INSPIRE as of 29 Jul 2019
  
\bibitem{W89}
  E.~Witten,
  ``Quantum Field Theory and the Jones Polynomial,''
  Commun.\ Math.\ Phys.\  {\bf 121}, 351 (1989).
  doi:10.1007/BF01217730
  %%CITATION = doi:10.1007/BF01217730;%%
  %2792 citations counted in INSPIRE as of 19 Jun 2019
\bibitem{HWZ}
  L.~Y.~Hung, Y.~S.~Wu and Y.~Zhou,
  ``Linking Entanglement and Discrete Anomaly,''
  JHEP {\bf 1805}, 008 (2018)
  doi:10.1007/JHEP05(2018)008
  [arXiv:1801.04538 [hep-th]].
  %%CITATION = doi:10.1007/JHEP05(2018)008;%%
  %4 citations counted in INSPIRE as of 16 Feb 2019
\bibitem{LS}
  Y.~H.~Lin and S.~H.~Shao,
  ``Anomalies and Bounds on Charged Operators,''
  arXiv:1904.04833 [hep-th].
  %%CITATION = ARXIV:1904.04833;%%
  %2 citations counted in INSPIRE as of 14 Jun 2019
\bibitem{JW}
  W.~Ji and X.~G.~Wen,
  ``Non-invertible anomalies and mapping-class-group transformation of anomalous partition functions,''
  arXiv:1905.13279 [cond-mat.str-el].
  %%CITATION = ARXIV:1905.13279;%%
\bibitem{NY}
  T.~Numasawa and S.~Yamaguch,
  ``Mixed global anomalies and boundary conformal field theories,''
  JHEP {\bf 1811}, 202 (2018)
  doi:10.1007/JHEP11(2018)202
  [arXiv:1712.09361 [hep-th]].
  %%CITATION = doi:10.1007/JHEP11(2018)202;%%
  %3 citations counted in INSPIRE as of 16 Feb 2019
\bibitem{HTHR}
  B.~Han, A.~Tiwari, C.~T.~Hsieh and S.~Ryu,
  ``Boundary conformal field theory and symmetry protected topological phases in $2+1$ dimensions,''
  Phys.\ Rev.\ B {\bf 96}, no. 12, 125105 (2017)
  doi:10.1103/PhysRevB.96.125105
  [arXiv:1704.01193 [cond-mat.str-el]].
  %%CITATION = doi:10.1103/PhysRevB.96.125105;%%
  %7 citations counted in INSPIRE as of 14 Jun 2019
\bibitem{Y19}
  Y.~Zhou,
  ``$3d$ One-form Mixed Anomaly and Entanglement Entropy,''
  arXiv:1904.06924 [hep-th].
  %%CITATION = ARXIV:1904.06924;%%
\bibitem{Verlinde:1988sn}
  E.~P.~Verlinde,
  ``Fusion Rules and Modular Transformations in 2D Conformal Field Theory,''
  Nucl.\ Phys.\ B {\bf 300}, 360 (1988).
  doi:10.1016/0550-3213(88)90603-7
  %%CITATION = doi:10.1016/0550-3213(88)90603-7;%%
  %875 citations counted in INSPIRE as of 17 Jun 2019 
\bibitem{Petkova:2000ip}
  V.~B.~Petkova and J.~B.~Zuber,
  ``Generalized twisted partition functions,''
  Phys.\ Lett.\ B {\bf 504}, 157 (2001)
  doi:10.1016/S0370-2693(01)00276-3
  [hep-th/0011021].
  %%CITATION = doi:10.1016/S0370-2693(01)00276-3;%%
  %132 citations counted in INSPIRE as of 17 Jun 2019
\bibitem{Fuchs:2002cm}
  J.~Fuchs, I.~Runkel and C.~Schweigert,
  ``TFT construction of RCFT correlators 1. Partition functions,''
  Nucl.\ Phys.\ B {\bf 646}, 353 (2002)
  doi:10.1016/S0550-3213(02)00744-7
  [hep-th/0204148].
  %%CITATION = doi:10.1016/S0550-3213(02)00744-7;%%
  %172 citations counted in INSPIRE as of 17 Jun 2019
\bibitem{Frohlich:2004ef}
  J.~Frohlich, J.~Fuchs, I.~Runkel and C.~Schweigert,
  ``Kramers-Wannier duality from conformal defects,''
  Phys.\ Rev.\ Lett.\  {\bf 93}, 070601 (2004)
  doi:10.1103/PhysRevLett.93.070601
  [cond-mat/0404051].
  %%CITATION = doi:10.1103/PhysRevLett.93.070601;%%
  %69 citations counted in INSPIRE as of 17 Jun 2019
\bibitem{Fuchs:2003id}
  J.~Fuchs, I.~Runkel and C.~Schweigert,
  ``TFT construction of RCFT correlators. 2. Unoriented world sheets,''
  Nucl.\ Phys.\ B {\bf 678}, 511 (2004)
  doi:10.1016/j.nuclphysb.2003.11.026
  [hep-th/0306164].
  %%CITATION = doi:10.1016/j.nuclphysb.2003.11.026;%%
  %54 citations counted in INSPIRE as of 17 Jun 2019
\bibitem{Fuchs:2004dz}
  J.~Fuchs, I.~Runkel and C.~Schweigert,
  ``TFT construction of RCFT correlators. 3. Simple currents,''
  Nucl.\ Phys.\ B {\bf 694}, 277 (2004)
  doi:10.1016/j.nuclphysb.2004.05.014
  [hep-th/0403157].
  %%CITATION = doi:10.1016/j.nuclphysb.2004.05.014;%%
  %70 citations counted in INSPIRE as of 17 Jun 2019
\bibitem{Fuchs:2004xi}
  J.~Fuchs, I.~Runkel and C.~Schweigert,
  ``TFT construction of RCFT correlators IV: Structure constants and correlation functions,''
  Nucl.\ Phys.\ B {\bf 715}, 539 (2005)
  doi:10.1016/j.nuclphysb.2005.03.018
  [hep-th/0412290].
  %%CITATION = doi:10.1016/j.nuclphysb.2005.03.018;%%
  %67 citations counted in INSPIRE as of 17 Jun 2019
\bibitem{Fjelstad:2005ua}
  J.~Fjelstad, J.~Fuchs, I.~Runkel and C.~Schweigert,
  ``TFT construction of RCFT correlators. V. Proof of modular invariance and factorisation,''
  Theor.\ Appl.\ Categor.\  {\bf 16}, 342 (2006)
  [hep-th/0503194].
  %%CITATION = HEP-TH/0503194;%%
  %53 citations counted in INSPIRE as of 17 Jun 2019
\bibitem{Frohlich:2006ch}
  J.~Frohlich, J.~Fuchs, I.~Runkel and C.~Schweigert,
  ``Duality and defects in rational conformal field theory,''
  Nucl.\ Phys.\ B {\bf 763}, 354 (2007)
  doi:10.1016/j.nuclphysb.2006.11.017
  [hep-th/0607247].
  %%CITATION = doi:10.1016/j.nuclphysb.2006.11.017;%%
  %115 citations counted in INSPIRE as of 17 Jun 2019
\bibitem{Runkel:2005qw} 
  I.~Runkel, J.~Fjelstad, J.~Fuchs and C.~Schweigert,
  ``Topological and conformal field theory as Frobenius algebras,''
  Contemp.\ Math.\  {\bf 431}, 225 (2007)
  [math/0512076 [math-ct]].
  %%CITATION = MATH/0512076;%%
  %22 citations counted in INSPIRE as of 17 Jun 2019
\bibitem{Davydov:2010rm}
  A.~Davydov, L.~Kong and I.~Runkel,
  ``Invertible Defects and Isomorphisms of Rational CFTs,''
  Adv.\ Theor.\ Math.\ Phys.\  {\bf 15}, no. 1, 43 (2011)
  doi:10.4310/ATMP.2011.v15.n1.a2
  [arXiv:1004.4725 [hep-th]].
  %%CITATION = doi:10.4310/ATMP.2011.v15.n1.a2;%%
  %18 citations counted in INSPIRE as of 17 Jun 2019
\bibitem{Frohlich:2009gb}
  J.~Frohlich, J.~Fuchs, I.~Runkel and C.~Schweigert,
  ``Defect lines, dualities, and generalised orbifolds,''
  doi:10.1142/9789814304634-0056
  arXiv:0909.5013 [math-ph].
  %%CITATION = doi:10.1142/9789814304634_0056;%%
  %40 citations counted in INSPIRE as of 17 Jun 2019
 \bibitem{CFT}
  P.~Di Francesco, P.~Mathieu and D.~Senechal,
  ``Conformal Field Theory,''
  doi:10.1007/978-1-4612-2256-9
  %%CITATION = doi:10.1007/978-1-4612-2256-9;%%
  %161 citations counted in INSPIRE as of 16 Feb 2019
\bibitem{Moore:1988qv}
  G.~W.~Moore and N.~Seiberg,
  ``Classical and Quantum Conformal Field Theory,''
  Commun.\ Math.\ Phys.\  {\bf 123}, 177 (1989).
  doi:10.1007/BF01238857
  %%CITATION = doi:10.1007/BF01238857;%%
  %658 citations counted in INSPIRE as of 17 Jun 2019
\bibitem{Moore:1989yh}
  G.~W.~Moore and N.~Seiberg,
  ``Taming the Conformal Zoo,''
  Phys.\ Lett.\ B {\bf 220}, 422 (1989).
  doi:10.1016/0370-2693(89)90897-6
  %%CITATION = doi:10.1016/0370-2693(89)90897-6;%%
  %430 citations counted in INSPIRE as of 17 Jun 2019
\bibitem{Moore:1989vd}
  G.~W.~Moore and N.~Seiberg,
  ``Lectures On Rcft,''
  RU-89-32, YCTP-P13-89.
  %%CITATION = RU-89-32, YCTP-P13-89;%%
  %9 citations counted in INSPIRE as of 17 Jun 2019
\bibitem{FFRS}
  J.~Frohlich, J.~Fuchs, I.~Runkel and C.~Schweigert,
  ``Duality and defects in rational conformal field theory,''
  Nucl.\ Phys.\ B {\bf 763}, 354 (2007)
  doi:10.1016/j.nuclphysb.2006.11.017
  [hep-th/0607247].
  %%CITATION = doi:10.1016/j.nuclphysb.2006.11.017;%%
  %114 citations counted in INSPIRE as of 25 Apr 2019
\bibitem{C89}
  J.~L.~Cardy,
  ``Boundary Conditions, Fusion Rules and the Verlinde Formula,''
  Nucl.\ Phys.\ B {\bf 324}, 581 (1989).
  doi:10.1016/0550-3213(89)90521-X
  %%CITATION = doi:10.1016/0550-3213(89)90521-X;%%
  %825 citations counted in INSPIRE as of 18 Jun 2019
\bibitem{Ishibashistate}
  N.~Ishibashi,
  ``The Boundary and Crosscap States in Conformal Field Theories,''
  Mod.\ Phys.\ Lett.\ A {\bf 4}, 251 (1989).
  doi:10.1142/S0217732389000320;
  %%CITATION = doi:10.1142/S0217732389000320;%%
  %328 citations counted in INSPIRE as of 18 Jun 2019
  T.~Onogi and N.~Ishibashi,
  ``Conformal Field Theories on Surfaces With Boundaries and Crosscaps,''
  Mod.\ Phys.\ Lett.\ A {\bf 4}, 161 (1989)
  Erratum: [Mod.\ Phys.\ Lett.\ A {\bf 4}, 885 (1989)].
  doi:10.1142/S0217732389000228
  %%CITATION = doi:10.1142/S0217732389000228;%%
  %58 citations counted in INSPIRE as of 18 Jun 2019
\bibitem{torsion}
  C.~Vafa,
  ``Modular Invariance and Discrete Torsion on Orbifolds,''
  Nucl.\ Phys.\ B {\bf 273}, 592 (1986).
  doi:10.1016/0550-3213(86)90379-2;
  %%CITATION = doi:10.1016/0550-3213(86)90379-2;%%
  %400 citations counted in INSPIRE as of 18 Jun 2019
  K.~S.~Narain, M.~H.~Sarmadi and C.~Vafa,
  ``Asymmetric Orbifolds,''
  Nucl.\ Phys.\ B {\bf 288}, 551 (1987).
  doi:10.1016/0550-3213(87)90228-8
  %%CITATION = doi:10.1016/0550-3213(87)90228-8;%%
  %454 citations counted in INSPIRE as of 18 Jun 2019
\bibitem{KK}
  K.~Kikuchi, ``Orbifolding $D_{2l}$ type WZW model," to appear.
\bibitem{TS}
  Y.~Tanizaki and T.~Sulejmanpasic,
  ``Anomaly and global inconsistency matching: $\theta$-angles, $SU(3)/U(1)^2$ nonlinear sigma model, $SU(3)$ chains and its generalizations,''
  Phys.\ Rev.\ B {\bf 98}, no. 11, 115126 (2018)
  doi:10.1103/PhysRevB.98.115126
  [arXiv:1805.11423 [cond-mat.str-el]].
  %%CITATION = doi:10.1103/PhysRevB.98.115126;%%
  %15 citations counted in INSPIRE as of 08 Mar 2019
\end{thebibliography}
\end{document}